\documentclass[12pt]{article}

\usepackage{epsfig,amsfonts}

\renewcommand{\Re}{\mathop{\rm Re}\nolimits}

\newcommand{\betac}{\beta_\mathrm{c}}
\newcommand{\tr}{\mathop{\mathrm{tr}}}
\newcommand{\HT}{\rm{\it {HT}}}
\newcommand{\HS}{\rm{\it {HS}}}

\def \be {\begin{equation}}
\def \ee {\end{equation}}                                         
\def \ba {\begin{eqnarray}}
\def \ea {\end{eqnarray}}

\begin{document}

\title{A study of the phase transition in
4D pure compact U(1) LGT on toroidal and spherical lattices.}

\author{Isabel Campos, Andr\'es Cruz and Alfonso Taranc\'on.}
\bigskip
\maketitle

\begin{center}
{\it Departamento de F\'{\i}sica Te\'orica, Facultad de Ciencias,\\
Universidad de Zaragoza, 50009 Zaragoza, Spain \\
\small e-mail: \tt isabel, tarancon, cruz@sol.unizar.es} \\
\end{center}
\bigskip

\begin{abstract}
We have performed a
systematic study of the phase transition in the pure 
compact U(1) lattice gauge theory in the extended coupling 
parameter space ($\beta , \gamma$) on toroidal and spherical lattices. 
The observation of 
a non-zero latent heat in both topologies for all  
investigated $\gamma \in [+0.2,-0.4]$,  
together with an exponent $\nu_{\rm {eff}} \sim 1/d$ when large enough
lattices are considered, lead us to conclude that the phase transition
is first order. For negative $\gamma$, our results point to an
increasingly weak first order transition as $\gamma$ is made more negative.

\end{abstract}

\newpage

\section{Introduction and overview}

In this work the 4$D$ pure compact U(1) gauge theory is revisited.
This model, being the prototype of gauge abelian interaction, has received
a lot of attention through the years, since it constitutes a preliminary step
in approaching QED. However, the questions related with the mechanisms
driving the transition and with the order of the phase transition itself,
have turned out to be a more controversial issue than one might expect
from its apparent simplicity.

Through this section we shall summarize the main results obtained up to
date.

The 4$D$ pure compact U(1) theory is known to undergo a phase transition
between a region in which the photons are confined (strong coupling)
and another region in which they carry a Coulomb potential 
(weak coupling). The theory can be discretized in the lattice using
the Wilson action
\begin{equation}
S_{\rm W} = \beta \sum_{\rm P} [1 - \Re \tr U_{\rm P}] \ .
\label{SW}
\end{equation}

Most of the work has been done on the hypercubic lattice with periodic
boundary conditions, namely on the hypertorus. 
On this topology the pioneer studies (see eg. \cite{CRE,LAU,BHA}) pointed to a
continuous transition since numerical simulations did not evidence
signals of metastability. The confinement-deconfinement transition
sets in at a coupling $\beta \sim 1$.
The largest lattice size simulated
in those works was $L=5$. However, as simulations in larger lattices
became available to computer resources, metastabilities were evidenced,
and two-peaked energy histograms could be observed from $L=6$ on (see eg.
\cite{JER,VIC,UBE}) pointing to a first order character
for the phase transition. This idea was also supported by several works
that approached the problem from the point of view of the Renormalization
Group \cite{REBBI2,PET,HAS,ALF}.

Nowadays, though the general belief is that the transition is 
first order, there are still some puzzling questions. Namely, the 
numerical simulations up to $L=12$ do not reveal a stable latent heat,
it is rather decreasing with the lattice size,
and also, the associated critical exponent $\nu$ is always in the
interval $\nu=0.29-0.32$, which is different from the first order value
($\nu=0.25$) and from the trivial second order one, namely $\nu=0.5$.
These facts led some authors to wonder about the possibility of a
slowly vanishing latent heat, which could eventually go to zero
in the thermodynamic limit.
In short, the 4$D$ pure compact U(1) theory could be critical
provided the two peaks appearing in the simulations are
a finite size effect. Due to the physical implications of a critical
theory with non trivial critical exponents
this possibility was not completely ruled out.

Now, we go back in time to introduce the extended Wilson action.
Besides the Wilson action, an action including a term proportional
to the square of the plaquettes was proposed by G. Bhanot at the 
beginning of the  eighties \cite{BHA2}
\begin{equation}
S_{\rm {EW}} = \beta \sum_{\rm p} [1 - \Re \tr U_{\rm p}] +
\gamma \sum_{\rm p} [1 - \Re \tr U_{\rm p}^2] \ .
\label{SEW}
\end{equation}

\begin{figure}[t]
\begin{center}
\epsfig{figure= 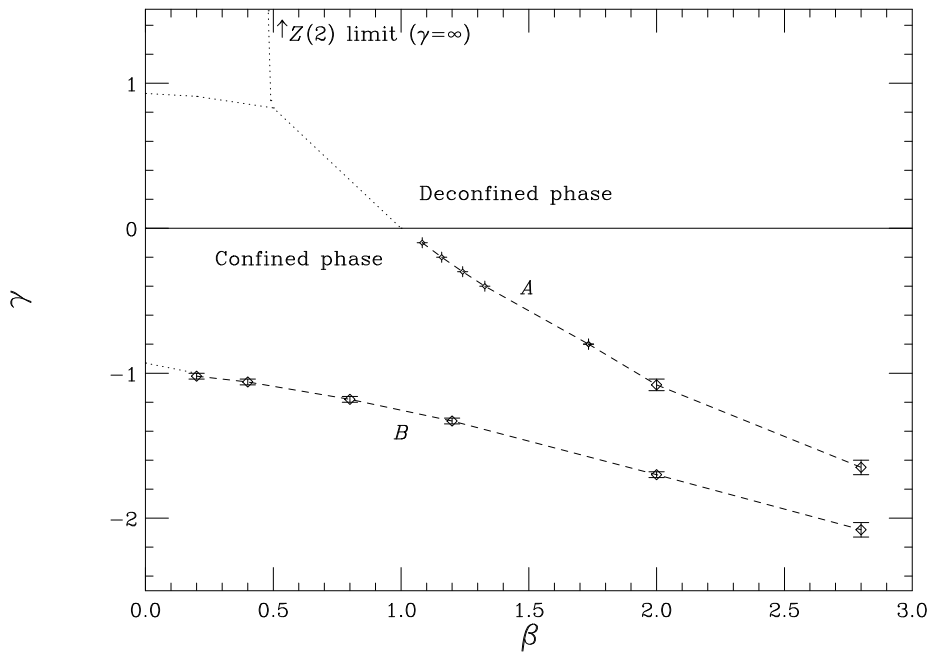,angle=90,width=350pt}
\caption{\small {Phase diagram of U(1) in the 
extended $(\beta, \gamma)$ parameter
space. The dotted lines have been taken from \protect\cite{BHA2}. 
The symbols correspond to our simulations on the torus. The crosses
correspond to the peak of $C_{\rm v}(L=16)$, the errors are not visible
in this scale. The diamonds have been obtained with hysteresis
cycles in $L=8$.}}
\label{PHD}
\end{center}
\end{figure}

In that work the extended ($\beta$,$\gamma$) parameter space was 
explored in $L=4,5$, and it was
suggested that the phase transition line $\betac(\gamma)$
became first order for $\gamma \geq +0.2$, while for $\gamma < 0.2$ the
transition appeared to be second order on those small lattices.
The phase diagram is represented in Figure \ref{PHD}. 

Simulations in larger hypercubic lattices have
shown that two metastable states are present for all, up to date, 
positive and negative simulated $\gamma$ values.
What holds true is the observation that 
a positive $\gamma$ value reinforces the transition
towards the first order, while a negative $\gamma$ value weakens the
transition. In fact, the latent heat, $C_{\rm {lat}}$, decreases
with $\gamma$, and then, a coupling where $C_{\rm {lat}}$ goes to zero
was supossed to exist at some finite (negative) value of $\gamma$.
This behavior induced some authors to conjecture about the existence of 
a tricritical point (TCP) at some negative coupling $\gamma_{\rm {TCP}}$
where the order of the phase transition
could change from being first order to be continuous \cite{TCP}.
However, every simulation performed at negative $\gamma$ has exhibited
double peak structures for large enough $L$. As for the Wilson action
($\gamma = 0$), for negative $\gamma$
a non stable latent heat is revealed by simulating in lattice
sizes in the range $L=4,12$.

From the point of view of the Renormalization Group this problem was
studied in \cite{ANA} and the results pointed to a phase transition
line $\betac(\gamma)$ of first order for all $\gamma$ except at the end,
$\gamma=-\infty$, where the transition is second order with trivial
exponents.

Nowadays, it is generally accepted that the transition is first order
for positive $\gamma$ values. However for $\gamma \leq 0$
two scenarios arise: 1) The phase transition
line $\betac(\gamma)$ is first order everywhere except at $\gamma = -\infty$;
2) There is a finite $\gamma_{TCP} \leq 0$ from which on, the phase transition
is second order.

As we have already remarked, all the previous discussions refer to
simulations performed on the hypertorus.

Recently some authors have cast a new hue on this problem. 
It was suggested some years ago that the jump in the energy observed when
the confining-deconfining transition takes place, could be due to the 
existence of closed monopole loops wrapping around the toroidal lattice,
which due to the particular topology cannot be contracted to a point
\cite{GUP}. This hypothesis led authors to work on lattices homotopic
to the sphere, since on this topology all monopole loops can be contracted
to a single point \cite{CN,BAIG,JCN}.

The relationship between the closed monopole loops and the discontinuity
in the energy is far from being proved. The influence of those non trivial
monopole loops on the phase transition has been studied by
Rebbi and coworkers \cite{REBBI3}. Their results suggest that 
the properties of the phase transition seem not to be
affected by the existence of such closed monopole loops.

However it is claimed the two state signal
does dissapear when working on a lattice constructed 
considering the surface of 
a 5$D$ cube, which has the topology of a sphere \cite{CN,JCN}. This single
peak structure together with the estimation of 
the critical exponent $\nu \sim 0.37$ measured at $\gamma = 0, -0.2, -0.5$,
has induced authors in \cite{JCN} to claim that the
double peak signal on the torus is a finite size effect, and
when working on the sphere one observes that indeed the
line $\betac(\gamma)$ for $\gamma \leq 0$ is a line of second order 
phase transitions,
all of them in the same universality class characterized by
a $\nu$ exponent around $1/3$. Assuming this second order scenario
for the pure gauge theory, fermions have been included (on the toroidal
topology) in order to study how fermions fit in the U(1) theory.
Also in this case a second order phase transition is conjectured
to exist \cite{COLMO}.

This situation is somehow disturbing, because
one would expect to get the asymptotic behavior of a phase transition
earlier on a lattice which is homogeneous and translationally invariant,
namely the hypertorus, than on a lattice which is manifestly non homogeneous,
 as is the surface of a 5D cube, and on which the translational 
invariance is only recovered in the limit of infinite volume.

First order phase transitions exhibit a wide variety of behaviors.
Some of them do not show any pretransitional effect, the correlation
length at the critical point, $\xi_c$, 
is small, and the thermodynamic quantities diverge as the volume
for reasonable lattice sizes. However when the correlation length is very
large, pretransitional effects take place, and they can also be very
significant. In literature these transitions are referred to as
Weak First Order (WFO) phase transitions and are characterized by
small discontinuities, such as a small latent heat.

Depending on whether or not
$L$ is larger than $\xi_c$, a WFO phase transition presents two differentiate
regions:

\begin{itemize}

\item{} {\sl Transient region} ($L \leq  \xi_c$). 
The transition behaves like
a second order one, since $\xi_c$ is effectively infinite. Thermodynamic
quantities diverge less than $L^d$ and hence the effective critical exponent
$\nu$ is larger than $1/d$.

\item{} {\sl Asymptotic region} ($L \gg \xi_c$). The first order character
of the transition is revealed by the onset of a stable two-peak structure
and an effective $\nu$ compatible with $1/d$.

\end{itemize}

How large should $\xi_c$ be for the transition to be considered  WFO 
seems a subjective matter.
Since $\xi_c \neq 0$ always, all first order transitions would be
weak in some extent. The key point for a transition to be considered
WFO is the existence of pretransitional effects. The pretransitional
behavior makes the transition along the transient region
to look like a second order one with
a definite critical coupling $\beta^{\prime}$. This second order phase
transition is preempted by the true first order one at the coupling $\betac$.
The distance between the couplings  $\beta^{\prime}$ and $\betac$
can be taken as a measure of the weakness of the transition \cite{LAF}.
Typical examples of such WFO transitions
are 2$D$ Potts models \cite{LAF} and the deconfining transition in finite
T QCD \cite{APE}. 

From this description it is clear that
the observation of no double-peak structures does not prove by itself 
the second order character of a phase transition. 

In view of this,
we decided to make a systematic study of the phase transition line
$\betac(\gamma)$ on the hyper-torus ($\HT$ topology hereafter)
and on the lattice with spherical
topology following \cite{CN,JCN} (which will be referred to as $\HS$
topology). On the hypertorus we improve on the
statistics running up to $L=24$ at some negative $\gamma$ values. On the sphere
we have run lattices substantially larger than those employed by authors
in \cite{JCN} to check if the two-state signal reappears when larger
spheres are considered. Our preliminary results, suggesting that the
double peak structure indeed reappears, are published somewhere
else \cite{PRIMER}. 
Here, the statistics on those lattice sizes have been increased,
and simulations in larger lattices are included. New observables
are used to extract our conclusions.

\section{Description of the model and observables}

We shall consider the parameter space described by the extended
Wilson action, which can be expressed in terms of the plaquette
angle in the following way:

\begin{equation}
S = - \beta \sum_{\rm p} \cos \theta_{\rm p} - \gamma \sum_{\rm p} 
\cos 2\theta_{\rm p}  \ .
\end{equation}

We use for the simulations the conventional $4D$ hypercube with
toroidal boundary conditions (hypertorus), and, for
comparison we also consider the surface of a $5D$ cube which is
topologically equivalent to a $4D$ sphere. Contrarily to the hypertorus,
the surface of a $5D$ cube is not homogeneous. There are a number of sites
which do not have the maximum connectivity, namely 8 neighbors. Due to this
fact, uncontrolled finite size effects are expected to turn up. 
Their influence can be somehow alleviated by the introduction of 
appropriate weight factors in those inhomogeneous sites \cite{JCN}.
Since the topology remains unchanged and we do not expect the order
of the phase transition to be affected by the rounding, we do not
use weight factors. 

As a notational remark, we label with $L$ and $N$ the side length
for the $\HT$ topology and for the $\HS$ topology respectively.

Next we define the energies associated to each term in the action
\begin{eqnarray}
E_{\rm p} = \frac{1}{N_{\rm p}} \langle \sum_{\rm p} \cos \theta_{\rm p} 
\rangle \ , \\
E_{\rm {2p}} = \frac{1}{N_{\rm p}} \langle \sum_{\rm p} \cos 2 \theta_{\rm p} 
\rangle  \ , 
\end{eqnarray}

where $N_{\rm p}$ stands for the number of plaquettes.
On the hyper-torus this number is simply proportional to the volume,
namely the forward plaquettes are
$N_{\rm p} = 6 L^4$. On the sphere the number of plaquettes has a
less simple expression, and can be computed as a function of $N$
as $N_{\rm p}= 60(N-1)^4+20(N-1)^2$. 
In this case, the system is
not homogeneous and $N_{\rm p}$ is not proportional to the number of points
on the four dimensional surface, which is $N^5-(N-2)^5$, some points having a 
number of surrounding plaquettes less than the possible maximum $12$, 
as opposed to what happens on the torus .
 In order to allow comparison, we 
define $L_{\rm {eff}}=(\frac{N_{\rm p}}{6})^{1/4}$, in such a way that a 
hypertorus of $L=L_{\rm {eff}}$ has the same number of plaquettes as
the corresponding hypersphere.

The specific heat and the Binder cumulant are useful quantities
to monitorize the properties of a phase transition. 
At the critical point, they are known to posses different thermodynamical 
limits depending on whether the transition is first order or higher order,
and hence, their behavior with increasing lattice size can give
some clues as to the order of the phase transition.
We have studied the Finite Size Scaling (FSS) of these two energy cumulants.

The specific heat is defined for both energies as:

\begin{equation}
C_{\rm v} = \frac{\partial}{\partial\beta}E_{\rm p} = 
N_{\rm p} ( \langle E_{\rm p}^2 \rangle - \langle E_{\rm p} \rangle^2 ) \ ,
\label{CSPE}
\end{equation}

\begin{equation}
C_{\rm {2v}} = \frac{\partial}{\partial \gamma}E_{\rm {2p}} = 
N_{\rm p} ( \langle E_{\rm {2p}}^2 \rangle - \langle E_{\rm {2p}} \rangle^2 ) \ .
\label{CSPE2}
\end{equation}

As we have observed a high correlation between both energies, the
results being qualitatively the same, we shall only report
on the observable defined for
the plaquette energy $E_{\rm p}$.

In a second order phase transition, scaling theory predicts the specific
heat maximum to diverge as $L^{\alpha/\nu}$.
If the transition is first order it is expected
to diverge like the volume $L^d$ (or strictly like the number of
plaquettes $N_{\rm p}$) reflecting that the maximum of the energy fluctuation
has the size of the volume. This is expected to hold in the asymptotic
region of the transition $L \gg \xi_c$. In the transient region,
namely $L < \xi_c$, the specific heat is expected to grow more slowly than
the volume. 

We shall use the number of plaquettes $N_{\rm p}$ to study the specific heat
behavior in order to have a single parameter for both the $\HT$ and 
the $\HS$ topologies. This is of course equivalent to doing the discussion
as a function of $L$ on the $\HT$ and $L_{\rm {eff}}$ on the $\HS$.
We have in this case $C_v \sim (N_{\rm p})^{\alpha/d\nu}$.

We have also studied the behavior of the fourth cumulant of the energy
\begin{equation}
V_L = 1 - \frac{\langle E_{\rm p}^4 \rangle_L}{\langle E_{\rm p}^2 
\rangle^2_L} \ .
\end{equation}

When the energy distribution describing the system is gaussian $V_L 
\rightarrow 2/3$ in the thermodynamic limit. This is the case for a second
order phase transition. If the transition is first order, far from
the critical coupling $\betac$, $V_L$ also tends to 2/3, reflecting the
gaussianity of the energy distribution. However, at $\betac$ the distribution 
can be described by two gaussians centered about the energy of each
metastable state $E_1$ and $E_2$, and hence this quantity has the
non-trivial thermodynamic limit 

\begin{equation}
V_{L \rightarrow \infty} \rightarrow 1 - \frac{2(E_1^4 + E_2^4)}{3(E_1^2 + E_2^2)^2} < 2/3  \ .
\end{equation}

Another interesting quantity measurable from the density of states
is the distribution of the partition function zeroes, or Lee-Yang
zeroes \cite{LEE}. To clarify the role of the partition function zeroes
let us first stress a well known fact: there are no phase transitions
on a finite volume, phase transitions arise in the thermodynamical limit
and are signaled by non-analyticity in the free energy:

\begin{equation}
F(\beta,V) = - \frac{1}{\beta V} \log Z(\beta) \ .
\end{equation}

$Z(\beta)$ is a linear combination of exponentials, and hence an analytical
function. This implies that the free energy can be singular only where
$Z(\beta) = 0$. However if the coupling $\beta$ is real, and the volume
is finite, that linear combination is a sum of positive terms and hence, the
zeroes of $Z(\beta)$ are located in the complex plane of the coupling $\beta$.
The onset of the phase transition in the limit of infinite volume
is signaled by a clustering of the zeroes on the real axis at $\betac$.

Here follows the description of our procedure, for a general description
of the method see \cite{ENZO}. 

As mentioned previously we work only with $E_{\rm p}$, and then we 
consider only the spectral density method \cite{FALCI} for this variable.
Since we work at fixed $\gamma$ we consider only the $\beta$ coupling
derivatives.
 
From the MC simulation at $\beta$ we obtain an approximation to the density
of states  which allows us to compute the normalized energy distribution
$P_{\beta} (E)$. The energy distribution can be expressed as

\begin{equation}
P_{\beta}(E) = \frac{1}{Z} W(E) e^{-\beta E} \ .
\label{PROB}
\end{equation}

We use the standard reweighting technique to obtain from the distribution
measured at $\beta$ the distribution 
at another coupling $\omega$, which is complex in the more general
case $\omega = \eta + i \xi$. The standard reweighting formula is:

\begin{equation}
P_{\omega}(E) = \frac{P_{\beta}(E) e^{-(\omega - \beta)E} }
{\sum_E P_{\beta}(E) e^{-(\omega - \beta)E} } \ ,
\end{equation}

using (\ref{PROB}) and the normalization condition $\sum_E P_{\beta}(E) =1$
one obtains

\begin{equation}
\frac{Z(\omega)}{Z(\beta)} = 
\sum_E P_{\beta}(E) e^{-(\omega - \beta)E} \ .
\end{equation}

We can factorize the contributions from the real and 
imaginary part:

\begin{equation}
\frac{Z(\omega)}{Z(\beta)} = 
\sum_E P_{\beta}(E) (\cos(E\xi) + i\sin(E\xi)) e^{-(\eta - \beta)E} \ .
\end{equation}

This is the standard reweighting formula extended to the complex parameter
space of couplings. As a first observation we have a pure oscillating factor
due to $Im(\omega) \neq 0$. Since $E$ is $O(V)$ this is a rapidly
oscillating function which makes it impossible to locate zeroes with
large imaginary part.

The real part of the coupling contributes to the well known
exponential damping, $e^{-(\eta - \beta)E}$, which is telling us that we can
trust the extrapolation only in a small neighborhood $\beta \pm \eta$.

Since $Z(\Re \omega)$ has no zeroes in a finite volume, 
an easy way to locate the zeroes numerically is looking at the minima
of the function  $|G(\omega)|^2$, where:

\begin{equation}
G(\omega) = \frac{Z(\omega)}{Z(\Re \omega)} = \frac
{\sum_E P_{\beta}(E) (\cos(E\xi) + i\sin(E\xi)) e^{-(\eta - \beta)E}}
{\sum_E P_{\beta}(E) e^{-(\eta - \beta)E}} \ .
\end{equation}

The function to minimize is:

\begin{equation}
 |G(\omega)|^2 = 
\frac{(\sum_E P_{\beta}(E) e^{-(\eta - \beta)E} \cos(E \xi))^2
 + (\sum_E P_{\beta}(E) e^{-(\eta - \beta)E} \sin(E \xi))^2 }
{(\sum_E P_{\beta}(E) e^{-(\eta - \beta)E})^2} 
\label{F2}
\end{equation}

One interesting property of the partition function zeroes concerns
the estimation of critical exponents. Denoting by $\omega_0$
the coupling where the first zero is located, the distance to the real axis
scales with the $\nu$ exponent:

\begin{equation}
Im(\omega_0) \sim L^{-1/\nu}   \ .
\label{NU}
\end{equation}

We shall use this property to compute the effective $\nu$ exponent.

From $P_{E} (\beta)_L$, we can measure the free energy gap, $\Delta F(L)$,  
which is the difference between 
the minima and the local maximum of the free energy \cite{FREE}.
We use the spectral density method to get, from the measured histograms,
a new histogram where both peaks have equal height. We take
the logarithm of those histograms and measure the energy gap.
A growing $\Delta F(L)$ in the asymptotic region of the transition
implies a first order phase transition. An increase proportional
to $L^{d-1}$ is expected \cite{FREE}.
For the transition to be
second order, $\Delta F(L)$ must stay constant with increasing lattice
sizes.

\section{Numerical Simulation}

\begin{table}[h]
{

\begin{center}
{
\small{
\begin{tabular}{|c|c||c|c|c|c|c|c|c|} \hline
\multicolumn{7}{|c|}{$\gamma = +0.2$} \\  \hline
$N$  &$L_{\rm {eff}}$ &$\beta_{\rm {sim}}$ &$\tau$ &$N_\tau$  
&$\beta^{\ast} (L)$ &${\rm Im}(\omega_0)$  \\ \hline
$4$  &$5.383$    &0.8910  &70   &29142  &0.8906(2) &0.0169(1)       \\ \hline
$5$  &$7.150$    &0.8855  &210   &8000  &0.8855(2) &0.00604(6)   \\ \hline
$6$  &$8.921$    &0.8834  &520   &3800  &0.88330(5) &0.00244(6)   \\ \hline
$7$  &$10.694$   &0.8818  &790   &2300  &0.88182(3) &0.00090(3)    \\ \hline
$8$  &$12.469$   &0.88095 &930   &1930  &0.88097(3) &0.00040(2)   \\ \hline
\hline \hline
\multicolumn{7}{|c|}{$\gamma = 0$} \\  \hline
$N$  &$L_{\rm {eff}}$ &$\beta_{\rm {sim}}$ &$\tau$ &$N_\tau$ 
&$\beta^{\ast} (L)$ &${\rm Im}(\omega_0)$  \\ \hline
$6$  &$8.921$    &1.0128 &240   &8300 &1.01340(5)  &0.00616(8)      \\ \hline
$8$  &$12.469$   &1.0125 &400   &2300 &1.0127(2)   &0.00196(4)     \\ \hline
$10$ &$16.021$   &1.0120 &780   &1200 &1.01212(3)  &0.00075(3)     \\ \hline
$12$ &$19.574$   &1.0119 &850   &1150 &1.01194(2)  &0.00031(2)     \\ \hline
$14$ &$23.123$   &1.0117 &920   &1900 &1.01168(2)  &0.00013(1)    \\ \hline
\hline \hline
\multicolumn{7}{|c|}{$\gamma = -0.2$} \\  \hline
$N$  &$L_{\rm {eff}}$ &$\beta_{\rm {sim}}$ &$\tau$&$N_\tau$   
&$\beta^{\ast} (L)$ &${\rm Im}(\omega_0)$  \\ \hline
$6$  &$8.921 $   &1.1587 &160   &2000  &1.1587(4)  &0.0107(3)   \\ \hline
$7$  &$10.694$   &1.1600 &320	&2800  &1.1597(1)  &0.00550(7)  \\ \hline
$8$  &$12.469$   &1.1597 &510   &1000  &1.1603(2)  &0.00321(4)   \\ \hline
$10$ &$16.021$   &1.1602 &680   &1500  &1.1604(2)  &0.00171(2)   \\ \hline
$12$ &$19.574$   &1.1604 &820   &1200  &1.1602(1)  &0.00083(1)   \\ \hline
$14$ &$23.123$   &1.1605 &900   &1100  &1.16048(5) &0.00044(1)   \\ \hline
$16$ &$26.684$   &1.1604 &1150  &1200  &1.16038(2) &0.00023(1)  \\ \hline
\end{tabular}
}
}
\end{center}

}
\caption[a]{\footnotesize{Statistics of the data obtained 
for the $\HS$ topology}}
\protect\label{TABLA_ESFERA}

\end{table}

\begin{table}[!t]
{

\begin{center}
{
\small{
\begin{tabular}{|c|c|c|c|c|c|} \hline
\multicolumn{6}{|c|}{$\gamma = -0.1$}   \\ \hline
$L$ & $\beta_{\rm {sim}}$ & $\tau$ & $N_{\tau}$ &$\beta^{\ast} (L)$ 
&$Im(\omega_0)$  \\ \hline \hline
$6$ &1.0720 &350   &1900 &1.0716(2) &0.0097(1)       \\ \hline
$8$ &1.0784  &640  &1400  &1.0786(2)  &1.1539(2)    \\ \hline
$12$ &1.0820 &820   &750  &1.0818(1) &0.00114(2)   \\ \hline
$16$ &1.08278 &930    &900   &1.0827(1)  &0.00040(2)   \\ \hline
$20$ &1.0833 &1150   &1100  &1.0833(1) &0.00020(1)    \\ \hline
\hline \hline
\multicolumn{6}{|c|}{$\gamma = -0.2$}   \\ \hline
$L$ & $\beta_{\rm {sim}}$ & $\tau$ & $N_{\tau}$ &$\beta^{\ast} (L)$ 
&$Im(\omega_0)$  \\ \hline \hline
$6$   &1.1460 &380   &2000     &1.1452(2)  &0.0115(1)            \\ \hline
$8$   &1.1535  &620   &1900    &1.1539(2)  &0.00506(6)           \\ \hline
$12$  &1.1582 &840   &1200     &1.1582(2)  &0.00152(3)           \\ \hline
$16$  &1.15935 &920   &900     &1.1593(1)  &0.00060(2)           \\ \hline
$20$  &1.1599 &1150 &900       &1.1599(1)  &0.00028(1)           \\ \hline
\hline \hline
\multicolumn{6}{|c|}{$\gamma = -0.3$} \\ \hline  
$L$&$\beta_{\rm {sim}}$ &$\tau$  &$N_{\tau}$ &$\beta^{\ast} (L)$ 
&$Im(\omega_0)$  \\  \hline \hline
$6$&1.2255   &340  &2000  &1.2237(4) &0.0138(1)     \\   \hline
$8$&1.2344   &560   &1800 &1.2340(1) &0.00623(4)     \\   \hline
$12$&1.2395   &770  &1000 &1.2395(2) &0.00190(3)    \\  \hline
$16$ &1.2410  &900  &900  &1.2410(1) &0.00084(2)    \\  \hline
$20$ &1.2416  &1100  &1200  &1.24156(5)  &0.00041(2)     \\  \hline
$24$ &1.2417  &1200   &1100 &1.24162(5)  &0.00022(1)     \\  \hline
\hline \hline
\multicolumn{6}{|c|}{$\gamma = -0.4$} \\ \hline
$L$&$\beta_{\rm {sim}}$ &$\tau$  &$N_{\tau}$ &$\beta^{\ast} (L)$ 
&$Im(\omega_0)$  \\ \hline \hline
$6$   &1.3090  &400  &1800   &1.3082(4)  &0.0152(1)          \\ \hline
$8$   &1.3192 &600  &1600    &1.3194(3)  &0.00704(5)           \\ \hline
$12$   &1.3258 &710  &1400   &1.3259(1)  &0.00238(3)            \\ \hline
$16$   &1.32775 &840  &900   &1.3278(1)  &0.00108(2)            \\ \hline
$20$   &1.3285 &930  &1600   &1.3284(1)  &0.00054(2)            \\ \hline
$24$   &1.3286 &1150  &1500  &1.3286(1)  &0.00029(1)             \\ \hline
\end{tabular}
}
}
\end{center}

}
\caption[a]{\footnotesize{Statistics of the data obtained
in the $\HT$ topology.}}
\protect\label{TABLA_TORO}

\end{table}

Most of the work has been done by simulating the subgroup 
$Z(1024) \subset U(1)$, since the phase transition associated to the discrete
group lies safely far away. The overrelax algorithm can be applied only
in the simulations with $\gamma = 0$ so that the gain in statistical
quality due to the overrelax effect is limited to the Wilson 
action, where we have simulated both, the full group $U(1)$
and the discrete one for the sake of comparison.
For $\gamma \neq 0$ we have simulated the discrete group since so the
simulation is considerably speeded up.

For every lattice size we consider, we perform trial runs to locate
the coupling $\beta^{\ast}(L)$ where the specific heat
shows a peak. We use the standard reweighting techniques to extrapolate
in a neighborhood of the simulated coupling.
Once the peak is located within an error in the fourth
digit of the coupling, we perform an intensive simulation there
to get $P_{E}(\beta^{\ast})_L$. The statistics performed at these
pseudo-critical
couplings are reported in Table \ref{TABLA_ESFERA} and Table \ref{TABLA_TORO}
for the $\HS$ and the $\HT$ topology respectively.

Typically we measure the energies every 10 $MC$ sweeps in order
to construct the energy histogram (\ref{PROB})
From this energy distribution we 
obtain the cumulants of the energy we are interested in, and the critical
exponents using Finite Size Scaling techniques.
We remark that one of the main sources of systematic error 
when measuring critical exponents, is the
indetermination in the coupling $\beta^{\ast}(L)$ where to measure.

The simulations have been done in the {\sl RTNN} machine consisting of 32
Pentium Pro 200MHz processors. The total CPU time used is the equivalent
of 6 Pentium Pro years.

We have updated using a standard Metropolis algorithm with 2 hits. The
acceptance has always been between 65\% and 75\%. 

In order to consider the statistical quality of the simulation,
following \cite{SOKAL} we define the unnormalized autocorrelation 
function for the energy
\be
C(t) = \frac{1}{N-t} \sum_{i=1}^{N-t} E_i E_{i+t} - \langle E \rangle^2 \ ,
\ee
as well as the normalized one
\be
\rho(t) = \frac{C(t)}{C(0)} \ .
\ee

The integrated autocorrelation time for the energy, $\tau^{int}$, can be 
measured using the window method
\be
\tau^{int}(t) = \frac{1}{2} + \sum_{t^{\prime} = 1}^t \rho(t^{\prime}) \ ,
\ee

for large enough $t$, which is in practice selected self-consistently.
We use $t$ in the range 5$\tau^{int}$, 10$\tau^{int}$, and we check
that the obtained $\tau^{int}$ remains stable as the window in $t$
is increased. 

We have always started from hot and cold configurations in order to make
sure that the system does not remain in a long living metastable state,
which could be interpreted as a Dirac sheet. The results coming from
both types of starts have always been indistinguishable.

\begin{figure}[h]
\begin{center}
\epsfig{figure= 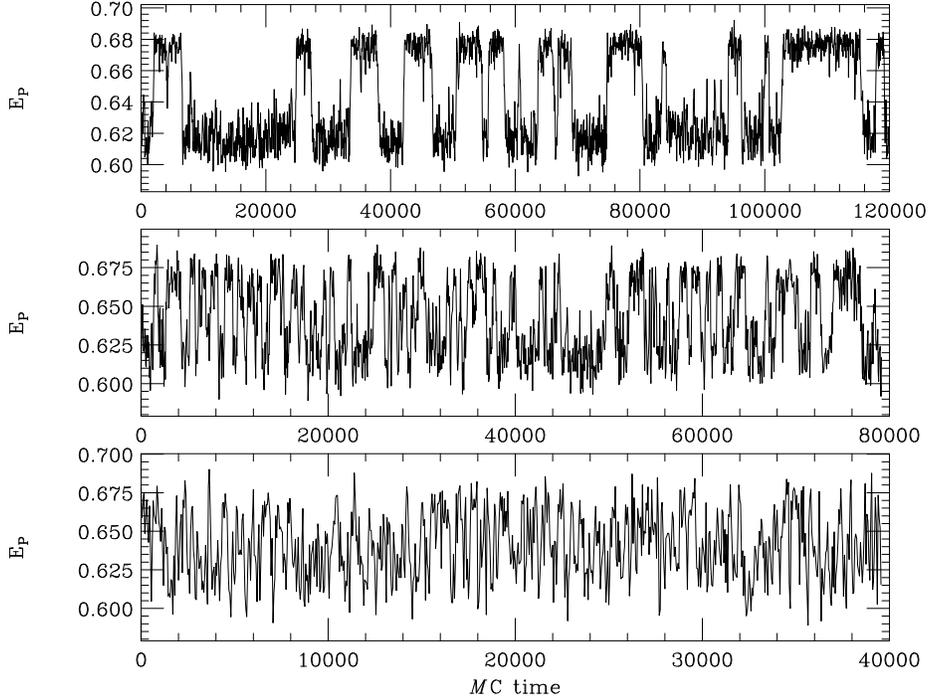,angle=90,width=350pt}
\caption{\small {MC evolution of $E_{\rm p}$ at $\gamma=+0.2$ on the $\HS$
topology for $N=6$ (lower window),$N=7$ (middle) and $N=8$ (top).}}
\label{EVOL02}
\end{center}
\end{figure}

\begin{figure}[h]
\begin{center}
\epsfig{figure= 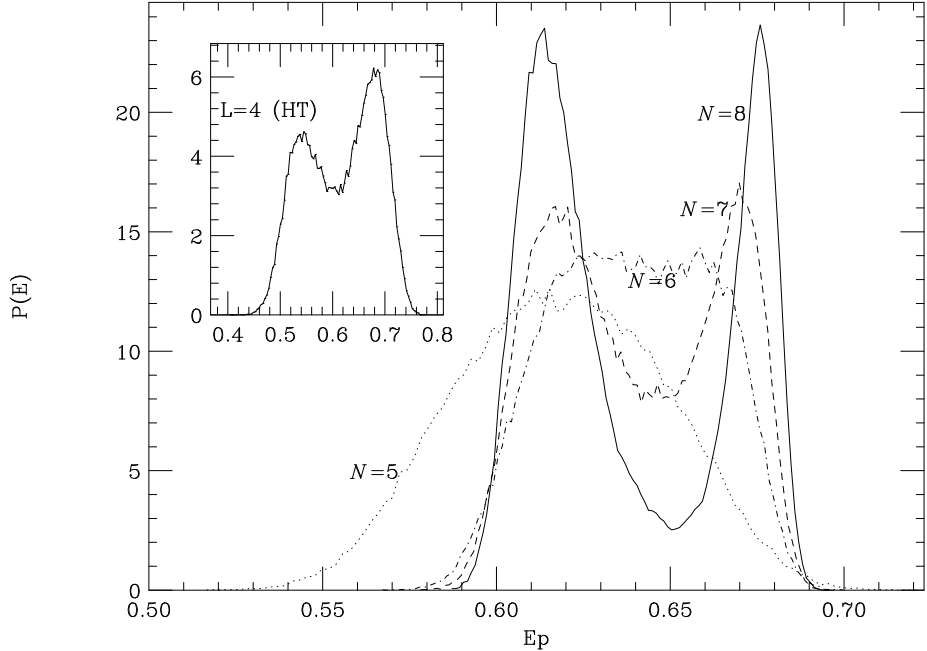,angle=90,width=350pt}
\caption{\small {$E_{\rm p}$ distribution at $\gamma=+0.2$ on the $\HS$
topology. The small window is the distribution we obtained on the
$\HT$ topology in $L=4$ at $\beta =0.8595$ }}
\label{HISTO02}
\end{center}
\end{figure}

\begin{figure}[h]
\begin{center}
\epsfig{figure= 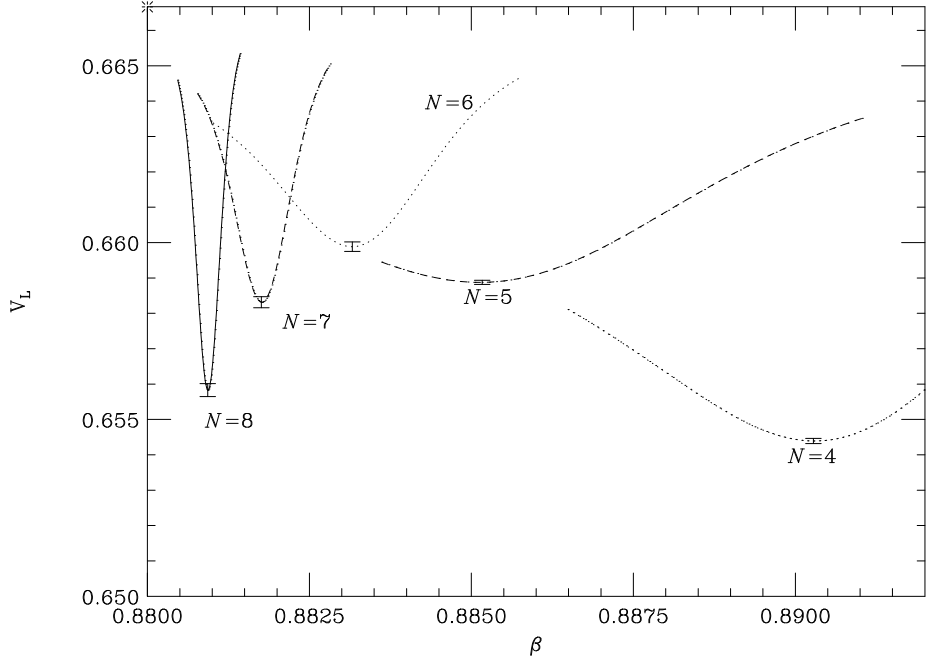,angle=90,width=350pt}
\caption{\small {Binder cumulant at $\gamma = +0.2$
on the $\HS$ topology in $N=4,5,6,7,8$. The cross in the upper corner 
signals the second order value $2/3$.}}
\label{BINDER02}
\end{center}
\end{figure}

\begin{figure}[h]
\begin{center}
\epsfig{figure= 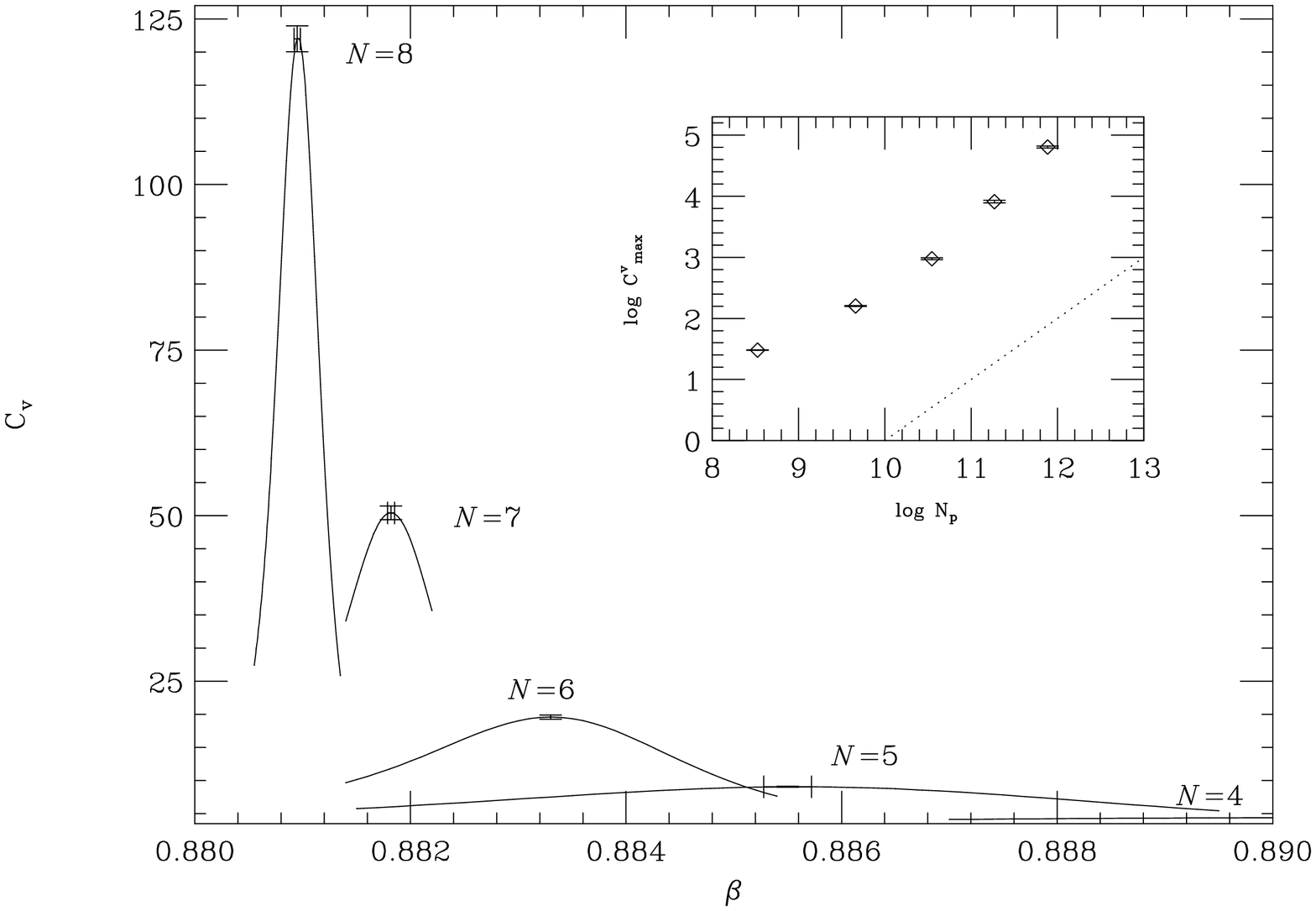,angle=90,width=350pt}
\caption{\small {Specific heat maximum, and Ferrenberg-Swendsen extrapolation 
(solid line) at $\gamma = +0.2$ on the $\HS$ topology.
The small window represents $C_{\rm v}^{\rm {max}} (N_{\rm p})$ 
The dotted line corresponds to the behavior expected in a
first order phase transition.}}
\label{PEAKS02}
\end{center}
\end{figure}

\section{The phase transition line $\betac(\gamma)$}

We have studied the deconfinement-confinement phase transition line
at several values of $\gamma$ (Figure \ref{PHD}).

In the region of negative $\gamma$, apart from the deconfinement
transition line, there is another transition line provoked by the
competing interaction between the couplings ($\beta,\gamma$).
In the limit $\gamma = -\infty$ the model is not dynamical since for
all finite $\beta$, $\cos \theta_{\rm p} = 0$. So we expect this
line to end at the corner $(\gamma = - \infty, \beta=+\infty)$.
We have not performed a deep study
of this transition line, however the simulations at $L=8$ revealed
double-peak structures pointing to a first order character.
The limit $\gamma = + \infty$ is equivalent to a $Z(2)$ theory and
the critical point can be calculated exactly by self-duality.

We have studied carefully the region
between lines $A$ and $B$
finding no signatures of the existence of additional lines.

We have focused on the transition line $A$,
at several values of $\gamma$, on the hyper-torus and on the hyper-sphere.

The structure of this section is the following:

First, we describe the results at $\gamma = +0.2$ on the spherical topology.
Our purpose is to observe how does the $\HS$ topology behave in a
non controversial region, and comparing with the known results on the
hypertorus.

Second, we study the point $\gamma = 0$ (Wilson action) on the spherical 
lattice. This coupling has been 
recently claimed to be the starting point of a critical line
(infinite correlation length at the critical point) which extends in
the range $\gamma \leq 0$, with an associated critical exponent
$\nu \sim 0.37$ \cite{JCN}. Since this assertion relies on the
absence of two-state signals on spherical lattices up to $N=10$,
we shall check whether or not double peak structures set in when
larger spheres are considered.

Third, we go to the region of negative $\gamma$. 
We are aware of no previous systematic study on the hypertorus in this region, 
so we run simulations up to $L=24$ at $\gamma = -0.1,-0.2,-0.3,-0.4$.
Motivated by the 
results on the hyper-torus we have just run a single $\gamma$
negative value on the hyper-sphere. For the sake of comparison with \cite{JCN}
we choose this value to be $\gamma = -0.2$.

Finally, we discuss the Finite Size Scaling behavior 
exhibited by both topologies.

\subsection{Results at $\gamma=+0.2$ on the spherical topology}

As we have pointed out,
there is a general agreement on considering the phase transition 
first order in the region of positive $\gamma$.

The lattice sizes used range from $N=4$ to $N=8$, which correspond
to $L_{\rm {eff}} \sim 5$ and $L_{\rm {eff}} \sim 12$ respectively
see (Table \ref{TABLA_ESFERA}). 

In Figure \ref{EVOL02}, the MC evolution for $N=6,7,8$ is shown. We remark that
no multicanonical update is needed to obtain a very high rate of flip-flops
up to $N=8$. However, on the torus, for $L > 6$ the probability of tunneling
between both metastable states is so tiny that a reasonable rate
of flip-flops is not accessible to ordinary algorithms.
Probably the inhomogeneity of the sphere
is increasing the number of configurations with energies which 
correspond neither to the confined nor to the deconfined phase, 
but in between.
These configurations make the free energy gap to decrease and hence
the tunneling is easier on the $\HS$ topology. In short, the sites without
maximum connectivity act as catalysts of the tunneling.

In Figure \ref{HISTO02} 
the energy distributions are plotted. The distribution at
$N=6$ is distinctly non-gaussian, but a blatant two-peak structure is
observed only from $N=7$ ($L_{\rm {eff}} \sim 11$) on. So, 
when comparing with the result at $L=4$ on the torus, (see 
small window in Figure \ref{HISTO02})
a first observation is that the 
onset of a two-state signal is particularly spoiled
by the $\HS$ topology, at least in 4$D$ pure compact U(1) gauge theory.

Concerning the latent heat,
we remark its stability already at $N=7$, or, if anything, its increase
from $N=7$ to $N=8$.
One would even say that the same happens
between the positions of the would-be two states in $N=6$ and the position
observed in $N=7$. The behavior of the Binder cumulant reflects this fact
(see Figure \ref{BINDER02}). We observe a rapid growth of $V_L^{\rm {min}}$
for small lattices sizes, apparently towards $2/3$ (second order value).
This growth stops when the splitting of the two peaks is observed.
The splitting of the two peaks is reflected by smaller values in
$V_L^{\rm {min}}$. We remark that the errors quoted for $V_L^{\rm {min}}$
are calculated taking into account the indetermination in the
value of the minimum, but not the displacement in the position of the
coupling where the minimum appears.

In our opinion, mainly two reasons can give account of this behavior. 
The first one comes from general grounds: at the very
asymptotic region of a first order phase transition, the 
energy jump gets sharper and sharper, and a slightly increase of the
latent heat could be expected. The second one would be
the increasing restoration of homogeneity in the hypersphere when
increasing the lattice size. The latent heat observed in small $N$
might be affected by the inhomogeneity of the hyper-sphere. 

At this point we cannot give a single reason for this to happen.
We postpone a stronger conclusion to the section devoted to the Finite
Size Scaling discussion.

The value of the latent heat when obtained with
a cubic spline fit to the peaks in $N=8$ is $C_{\rm {lat}} = 0.064(2)$.
Results obtained with mixed hot-cold starts in $N=9$ seems to give
an energy jump around 0.067. However the reliability of this
method is very limited and we do not dare to extract strong conclusions 
from it.
Taking into account that he cubic spline at $N=7$ gives 0.053(2) a possible
scenario would be a slowly increasing  latent heat towards its
asymptotic value $C_{\rm {lat}}(\infty)$.

The peak of the specific heat for the different lattice sizes is displayed
in Figure \ref{PEAKS02}.
The continuous line represents the $FS$ extrapolation.
The plot of the peak value, $C_v^{\rm {max}}(N_{\rm p})$ as a function of the
plaquette number reveals a linear relationship, and hence a first order
character.

\subsection{Results at $\gamma =0$ on the spherical topology}

\begin{figure}[!b]
\begin{center}
\epsfig{figure= 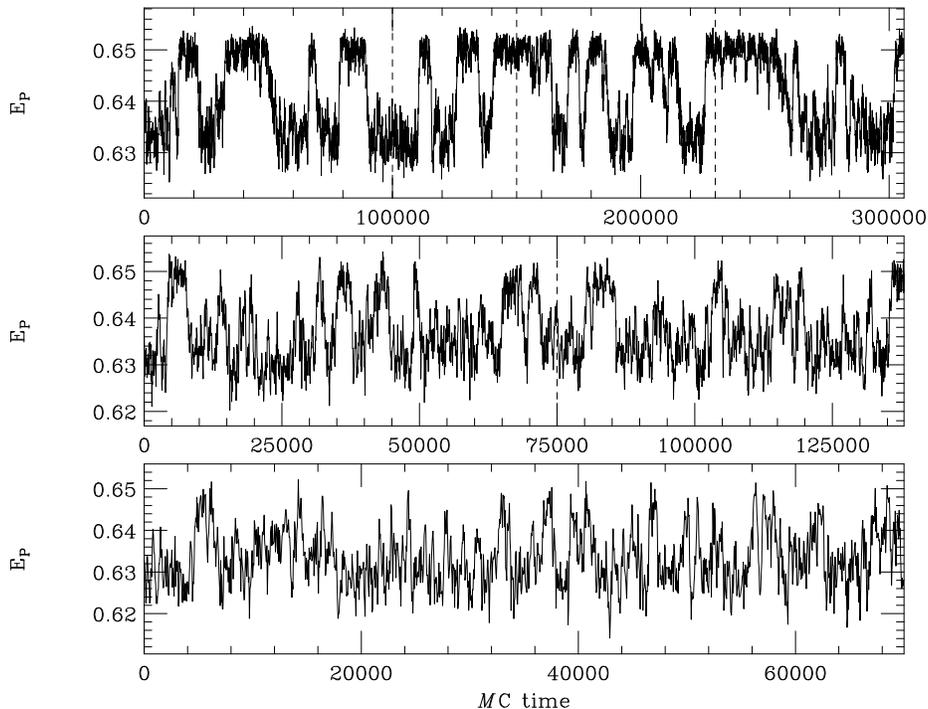,angle=90,width=350pt}
\caption{\small {MC evolution of $E_{\rm p}$ at $\gamma=0$ on the $\HS$
topology for $N=10$ (lower window), $N=12$ (middle) and $N=14$ (top).
The vertical dashed lines separate different runs}}
\label{EVOL0}
\end{center}
\end{figure}

\begin{figure}[h]
\begin{center}
\epsfig{figure= 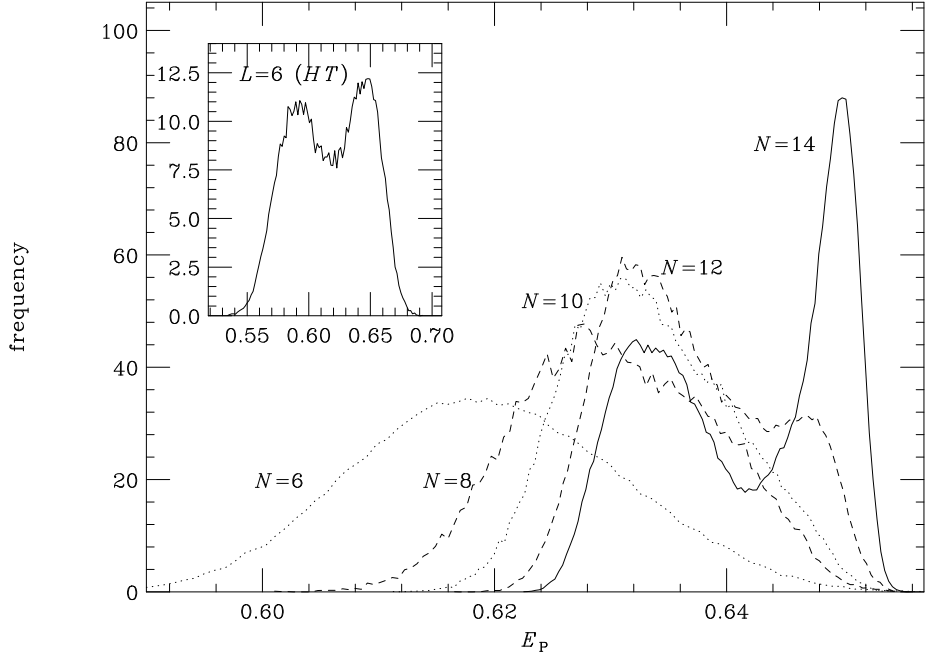,angle=90,width=350pt}
\caption{\small {$E_{\rm p}$ distribution at $\gamma=0$ on the $\HS$
topology. The small window corresponds to our simulation in the $\HT$
topology in L=6 at $\beta = 1.0020$.}}
\label{HISTO0}
\end{center}
\end{figure}

\begin{figure}[h]
\begin{center}
\epsfig{figure= 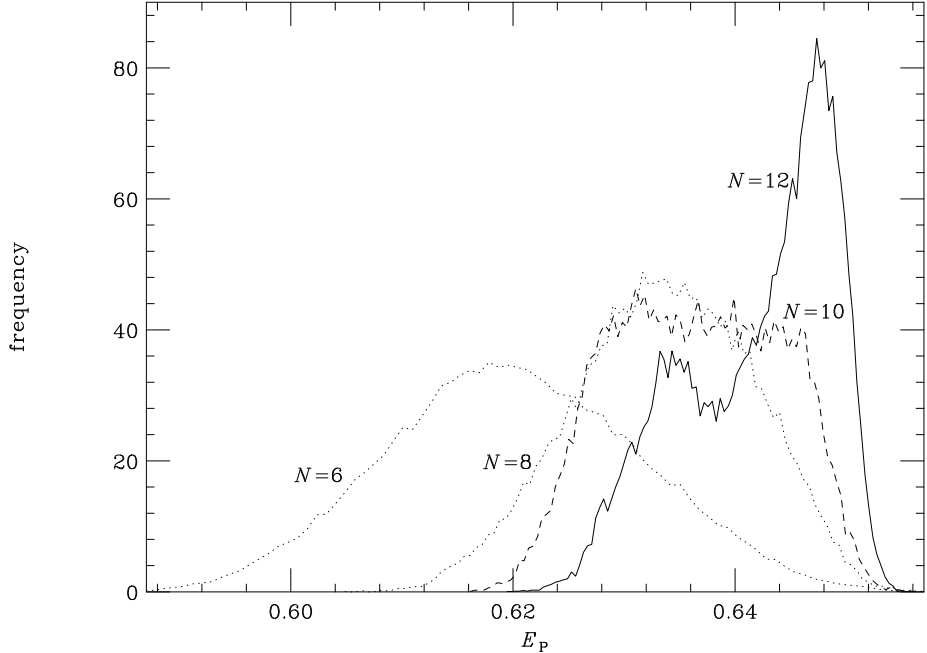,angle=90,width=350pt}
\caption{\small {$E_{\rm p}$ distribution at $\gamma=0$ on the $\HS$
topology simulating the full group U(1).}}
\label{HISTO0O}
\end{center}
\end{figure}

\begin{figure}[h]
\begin{center}
\epsfig{figure= 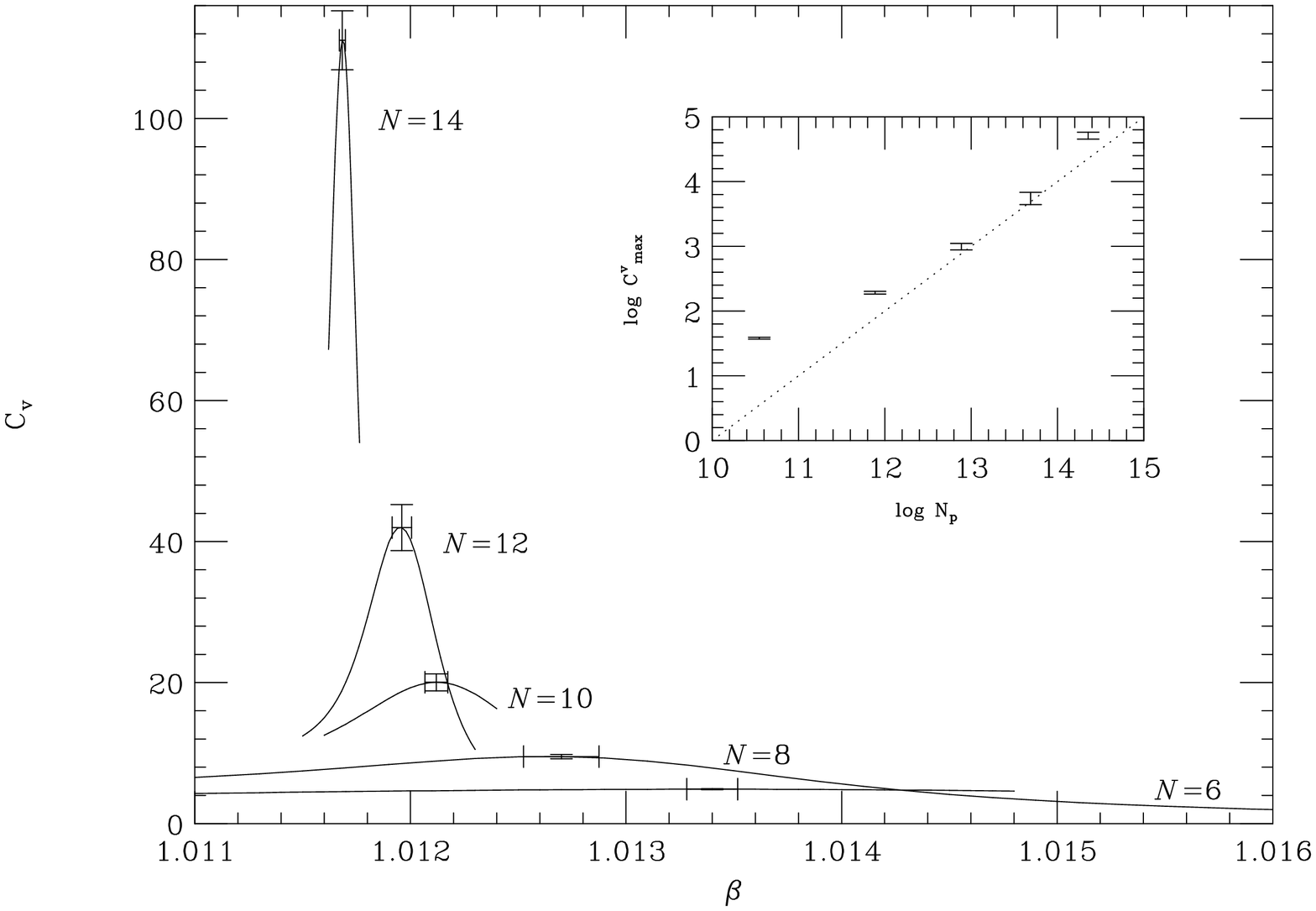,angle=90,width=350pt}
\caption{\small {Specific heat maximum, and Ferrenberg-Swendsen extrapolation 
(solid line) at $\gamma = 0$ on the $\HS$ topology. 
The small window represents $C_{\rm v}^{\rm {max}} (N_{\rm p})$ 
The dotted line corresponds to the behavior expected in a
first order phase transition. }}
\label{PEAKS0}
\end{center}
\end{figure}

\begin{figure}[h]
\begin{center}
\epsfig{figure= 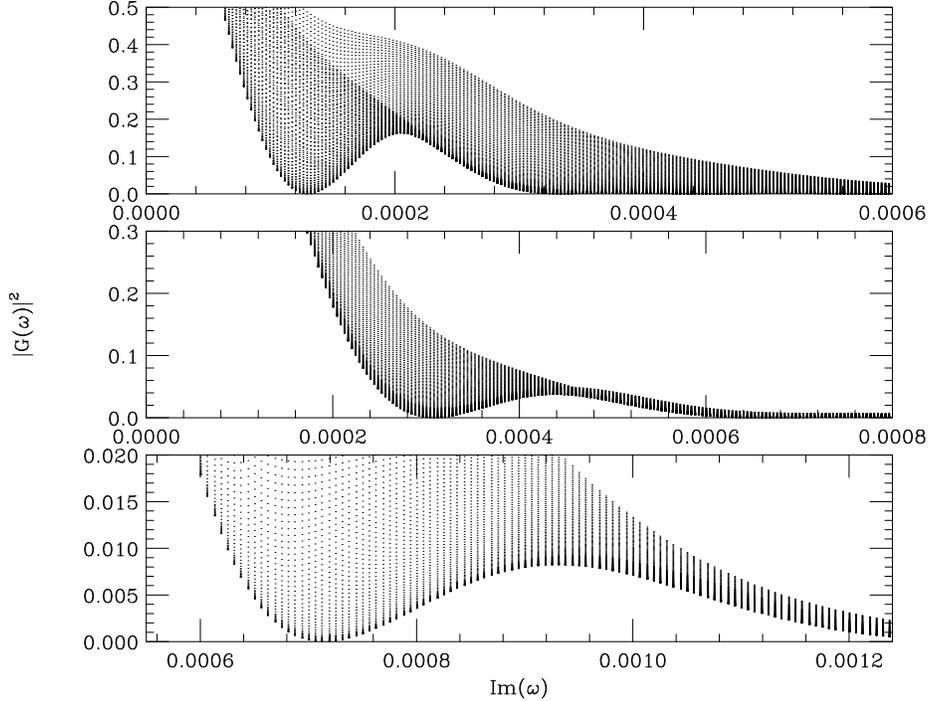,angle=90,width=350pt}
\caption{\small {Plot of (\protect\ref{F2}) for $N=10$ (lower window), 
$N=12$ (middle) and $N=14$ (top) at $\gamma = 0$ in the $\HS$ topology.}}
\label{FISHER0}
\end{center}
\end{figure}

\begin{figure}[h]
\begin{center}
\epsfig{figure= 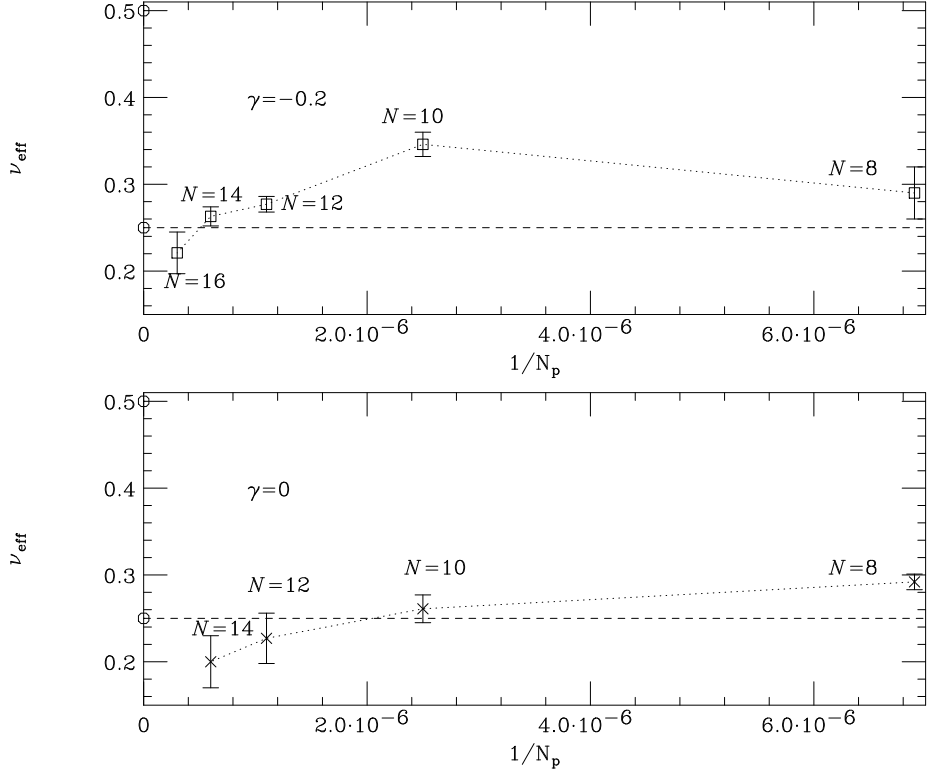,angle=90,width=350pt}
\caption{\small {Effective exponent $\nu$ at $\gamma = 0$ 
(lower window), and at
$\gamma = -0.2$ (upper window) on the spherical topology. In the smaller
lattices an $\nu_{\rm {eff}} \sim 1/3$ is observed which becomes $1/d$
when large enough lattices are considered.}}
\label{NUESFERA}
\end{center}
\end{figure}

\begin{figure}[h]
\begin{center}
\epsfig{figure= 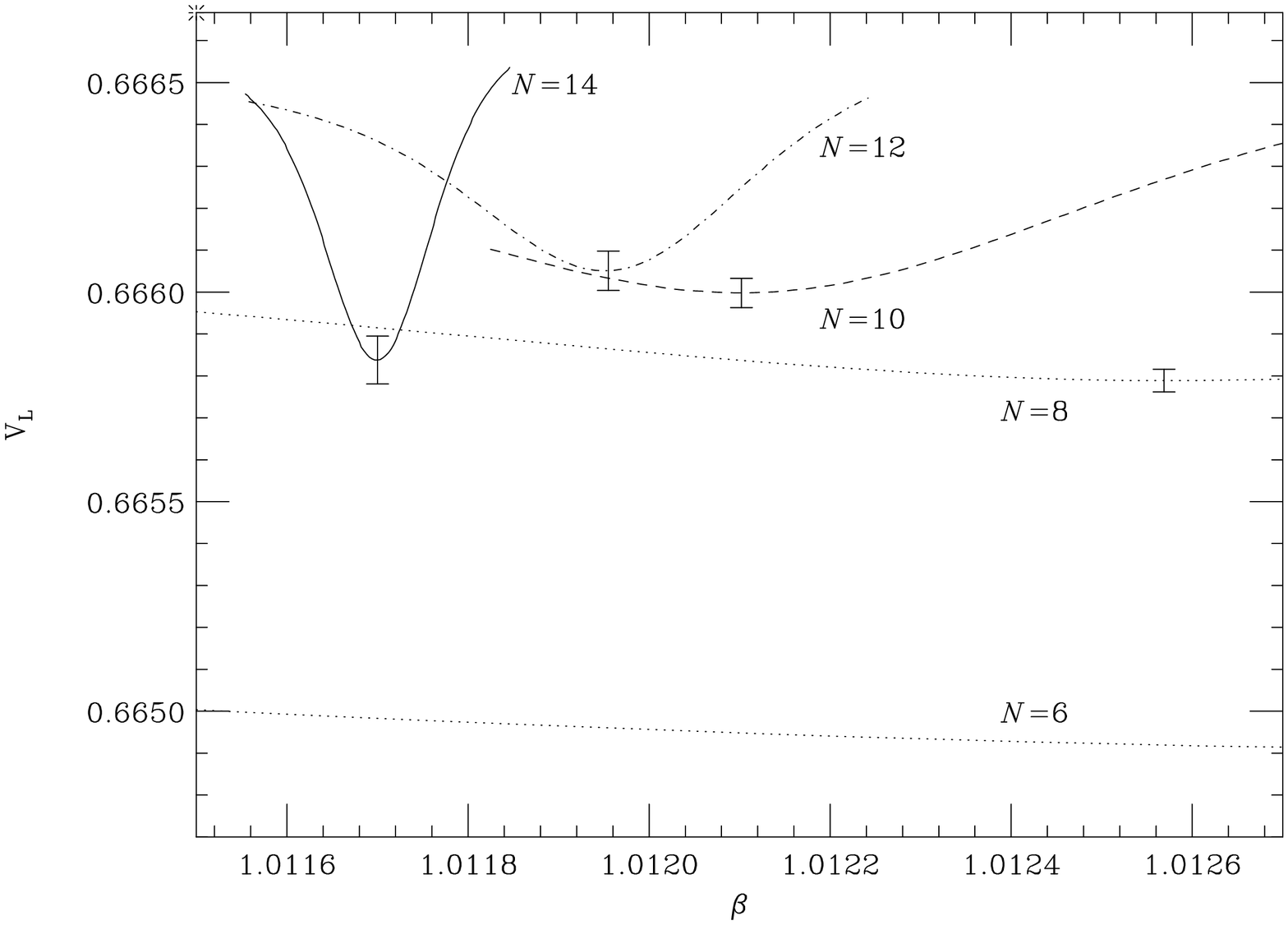,angle=90,width=350pt}
\caption{\small {Binder cumulant at $\gamma = 0$ (Wilson action)
on the $\HS$ topology in $N=7,8,10,12,14$. The cross in the upper corner 
signals the second order value $2/3$.}}
\label{BINDER0}
\end{center}
\end{figure}

Let us first situate the status of the studies using the toroidal topology
with the Wilson action.
As we have pointed out in the introduction, the transition was believed
to be continuous till the simulation of a $L=6$ lattice revealed
the first signs of the existence of two metastable states \cite{JER}. 
Numerical simulations up to $L=10$ showed a rapid decrease of the latent
heat when increasing the lattice size, suggesting that the two metastable
states might superimpose in the limit $L \rightarrow \infty$ \cite{VIC}.
However, in \cite{VIC2}, Azcoiti et al. suggest that the latent heat
starts to stabilize at $L \sim 12,14$. 

When one tries to simulate large lattices the tunneling becomes scarce and
the use of multicanonical simulations for those lattice sizes seems to be
in order. However, in spite of its usefulness in spin models, we are
aware of no multicanonical simulation showing a substancial flip-flop
rate improvement for this model. An alternative procedure to enhance
the tunneling probability has been proposed in \cite{REBBI}. Within this
approach the use of a monopole term in the action as a dynamical variable,
has been revealed as a useful method to increase the flip-flop rate.

Lattice sizes up to $L=16$ have also been studied using multihistogramming
techniques \cite{UBE} and RG approaches \cite{HAS,ALF}. The results
support the idea of a quasi stable latent heat for $L > 12$.

Topological considerations stressed already
in the introduction led some authors to use lattices homotopic 
to the sphere.
In the case of the Wilson action ($\gamma = 0$) the two-state signal
is absent up to $N=10$ \cite{JCN}. Following this, we have simulated
on the spherical topology in lattices ranging from $N=6$ to $N=14$,
finding that the two-state signal sets in from $N=12$ on. The preliminary
results up to $N=12$ are already published \cite{PRIMER}. We include here
the results with higher statistics in $N=12$ and the results in
$N=14$.

The MC evolution for $N=10,12,14$ is plotted in Figure \ref{EVOL0}.
For the $N=14$ lattice we have four independent runs, signaled
by the dashed lines in the figure, all of them giving the same 
predictions.

In Figure \ref{HISTO0} the distribution of $E_{\rm p}$ is plotted in lattices
ranging from $N=6$ to $N=14$. We observe that a two-peak structure is
revealed first time by the histogram in $N=12$, which has a 
$L_{\rm {eff}} \sim 19$. On the toroidal topology the equivalent 
signal is observed already at $L=6$ (see small window in Figure \ref{HISTO0}).

Having in mind the results of the previous section at $\gamma = +0.2$,
it seems that the minimum $L_{\rm {eff}}$ required 
to observe a two-peak structure
in the spherical topology is around three times the minimum lattice size $L$
needed in the torus to observe two peaks. 

The lattice sizes used in \cite{JCN} at $\gamma =0 $
ranged from $N=4$ to $N=10$, so it is not surprising that they did not see
any two-peak structure. Our results are in that sense compatible with
theirs, though ours show a faster divergence
for $C_{\rm v}^{\rm {max}}(N)$ with increasing $N$ than the one
observed in \cite{JCN} in the lattice sizes we share (N=6,8,10).
However we are particularly confident on this point because our
simulation has been performed closer to the peak of the specific
heat $\beta^{\ast}(L)$, and so we expect the Ferrenberg-Swendsen
extrapolation to be more precise.

For the sake of comparison with the full group U(1),
simulations with the Wilson action have been performed at $\gamma=0$.
At this coupling an implementation of overrelax is possible,
and the global decorrelating effect should manifest itself in a better
statistical quality.

In Figure \ref{HISTO0O} we show the histograms of $E_{\rm p}$ from
$N=6$ to $N=12$ simulating the full group. The results are
fully compatible with those obtained from simulations with $Z(1024)$.

This test being performed, we go back to the description of
the results obtained for the discrete group.

In Figure \ref{PEAKS0} we plot $C_{\rm v}(N)$ for different $N^{\prime}$s,
together with the Ferrenberg-Swendsen extrapolation. The small
window is a $\log - \log$ plot of $C_v^{\rm {max}}(N)$ as a function
of $N_{\rm p}$. 
A linear behavior is observed from $N=10$ on, which is even faster
than linear when $N=14$ is taken into account. This fact by itself implies
the first order character of the transition since it means that the
maximum in the energy fluctuation has the size of the volume.
As a further check we have measured the $\nu$ exponent from the scaling
of the Fisher zeroes (see Table \ref{TABLA_ESFERA}).

In Figure \ref{FISHER0} we plot $|G(\omega)|^2$ for several lattice
sizes on the spherical topology at $\gamma=0$. The different curves
stand for the different $\omega$ we extrapolate.
For small lattice sizes $Im(\omega_0)$ is larger, 
and the damping is more severe
than for the larger lattices since in the later the imaginary part
contributes with a faster oscillating function. Actually, we observe
that for $N=14$ even a second minimum could be measured before the signal
is damped, while for $N=10$ one can measure accurately only the first
one.

The results for the
effective $\nu$ are plotted in Figure \ref{NUESFERA} lower window.
As could be expected from the behavior of the specific heat, for small
lattices the effective exponent is somewhat larger than $0.25$. It
gets compatible with the first order value from $N=10$ on. In the largest
lattices the distance between the two peaks slightly increases, and we
measure a $\nu_{\rm {eff}}$ slightly smaller than the first order value.
As a $\nu < 0.25$ is strictly impossible we expect this to be a transient
effect due to finite size effects associated to the observed splitting 
up of the two peaks.

Concerning the latent heat, we observe a behavior completely analogous
to the one observed at $\gamma = +0.2$. We do observe a two
peak structure quite stable when comparing $N=12$ and $N=14$,
but it slightly increases in $N=14$. The plot of the Binder cumulant
reflects again this fact (see Figure \ref{BINDER0}). The fast growth
of $V_L^{\rm {min}}$ towards $2/3$ is preempted by the onset of
double peaked histograms from $N=12$ on, and it even decreases in $N=14$.

The cubic spline fit in $N=14$ at the peaks gives for
the latent heat $C_{\rm {lat}} \approx 0.018(2)$.
which is compatible with the results suggested by extrapolating
to infinite volume the values obtained on the torus up to $L=14$. 
Our results are hence supporting the conjecture stressed in \cite{VIC2}
about a quasi stabilization of the latent heat on the torus
from $L=12$ on.

\begin{figure}[h]
\begin{center}
\epsfig{figure= 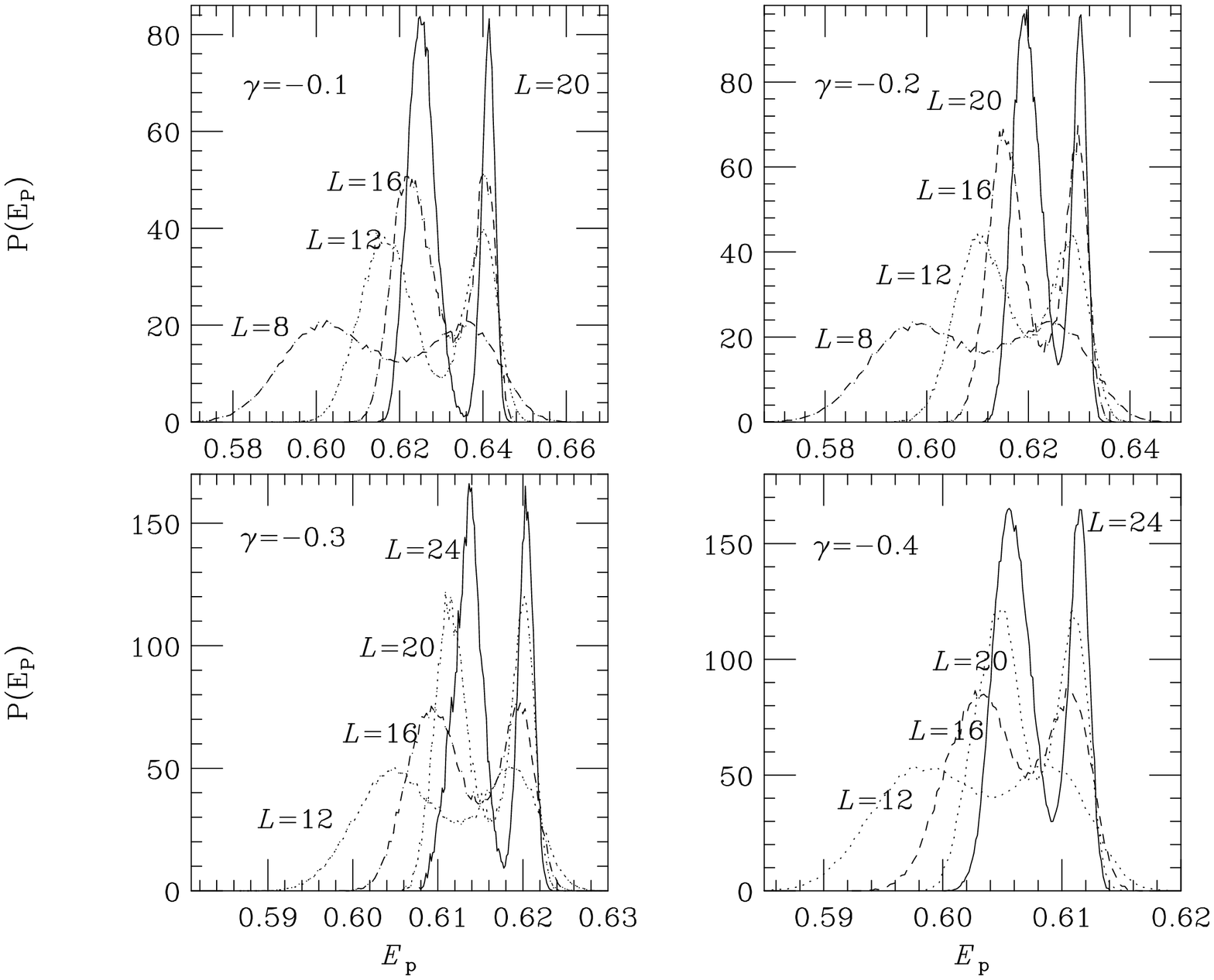,angle=90,width=350pt}
\caption{\small {$E_{\rm p}$ distributions at the different 
negative $\gamma$ on the
$\HT$ topology. We have run up to $L=20$ at $\gamma=-0.1,-0.2$ and up to
$L=24$ at $\gamma=-0.3,-0.4$.}}
\label{HISTO_TORO}
\end{center}
\end{figure}

\begin{figure}[h]
\begin{center}
\epsfig{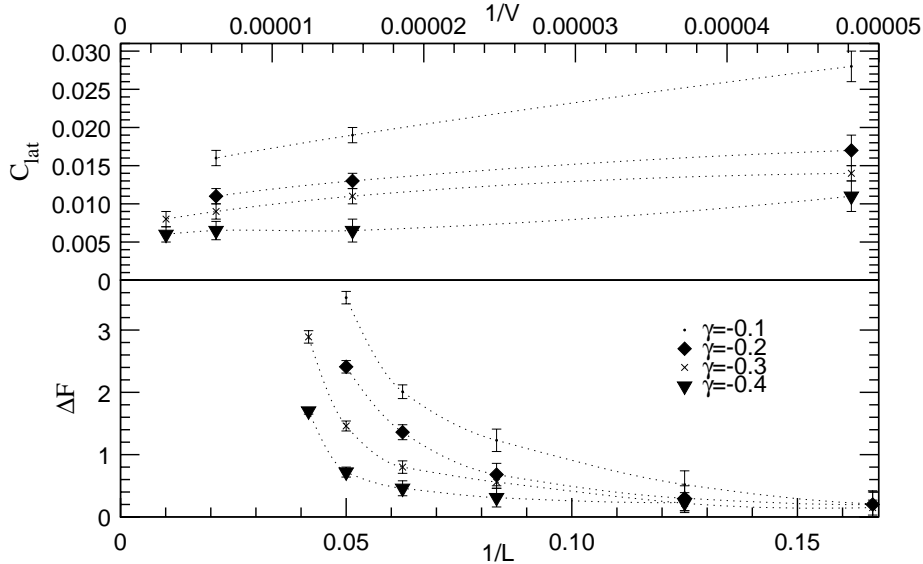}
\caption{\small {Free energy gap $\Delta F(L)$ (lower window) 
and latent heat (upper window) for the different negative $\gamma$ in 
the $\HT$ topology.}}
\label{GAP_LAT}
\end{center}
\end{figure}

\begin{figure}[h]
\begin{center}
\epsfig{figure= 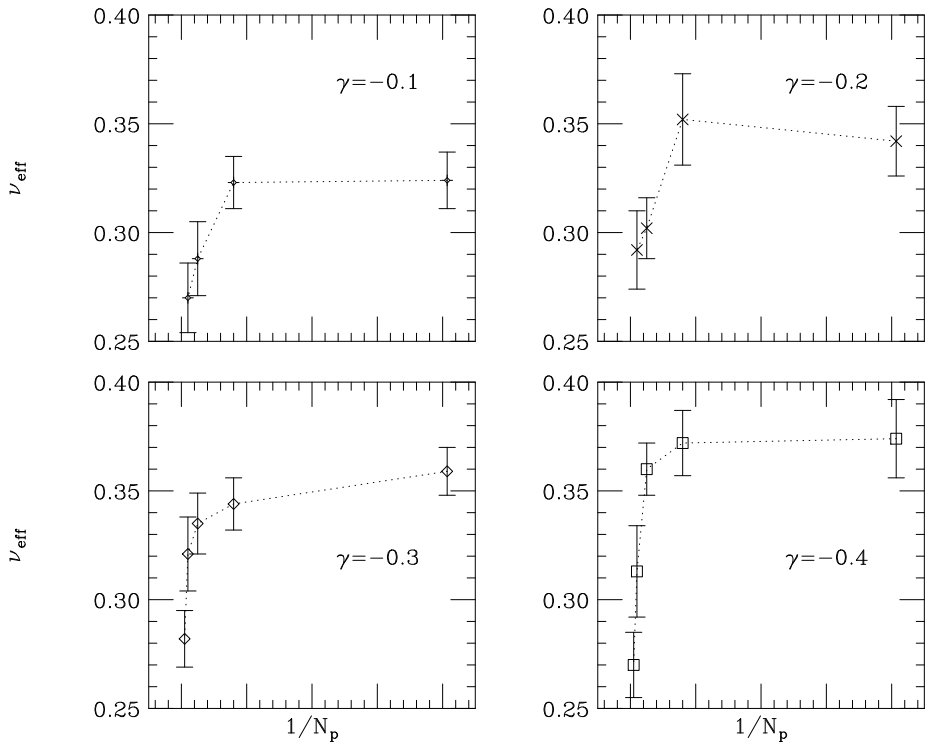,angle=90,width=350pt}
\caption{\small {Effective $\nu$ exponent for the different 
negative $\gamma$ in the $\HT$ topology.}}
\label{NU_TORO}
\end{center}
\end{figure}

\begin{figure}[h]
\begin{center}
\epsfig{figure= 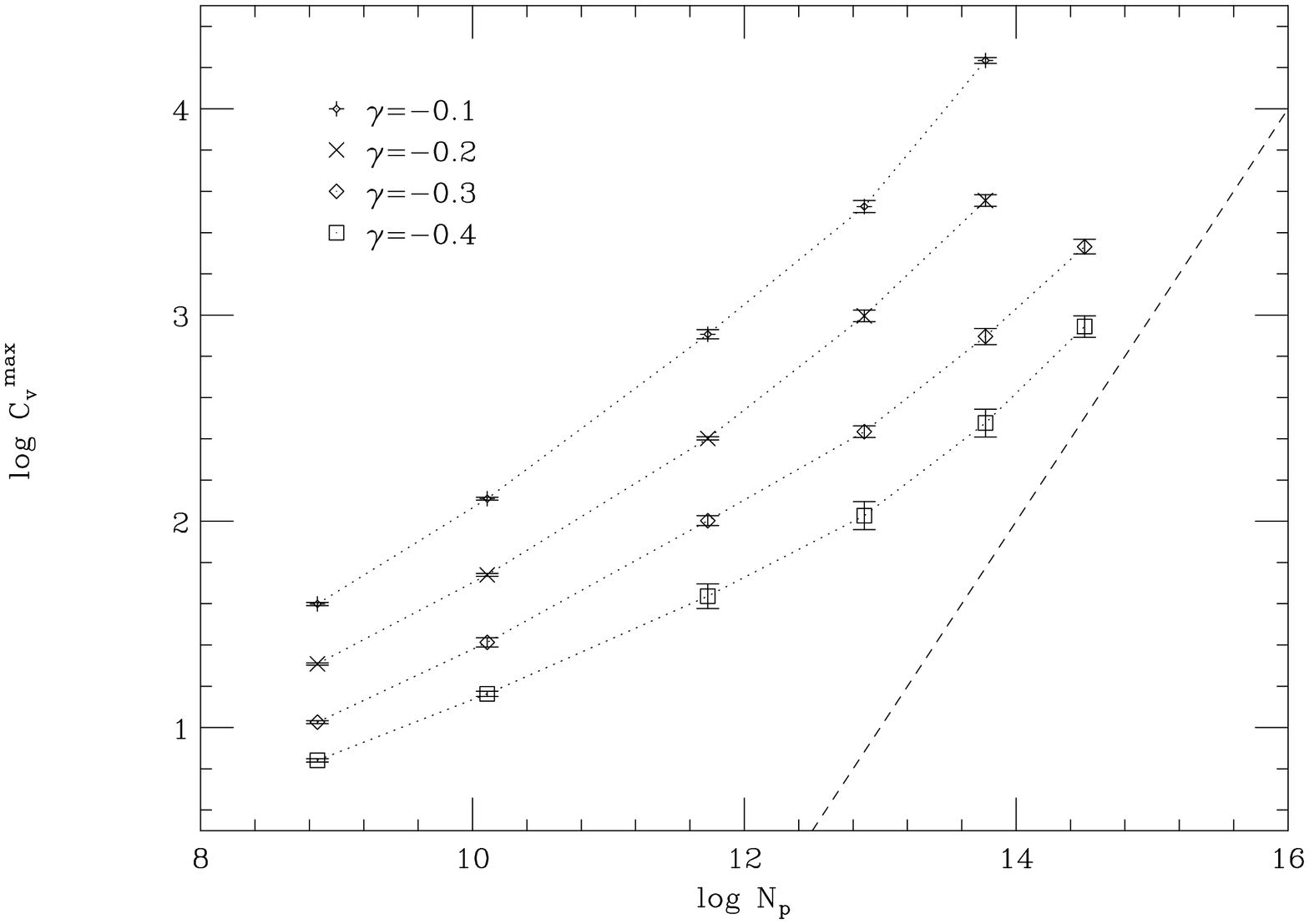,angle=90,width=350pt}
\caption{\small {$C_v^{\rm {max}}(L)$ as a function of $N_{\rm p}$ 
for the different negative $\gamma$ values on the $\HT$ topology. 
We have used $\log - \log$ scale 
for the clarity of the
graphic sake. The dotted line corresponds to the first order behavior.}}
\label{PEAKS_TORO}
\end{center}
\end{figure}

\subsection{Results for $\gamma < 0$}

\subsubsection{Toroidal topology}

We have performed a systematic study of the transition at several
negative values of $\gamma$ on the toroidal topology.  

We find that the two-state signal persists for all $\gamma$ values
we consider. The energy histograms reveal an increasing weakness
of the transition when going to more negative $\gamma$ values
(see Figure \ref{HISTO_TORO}). A double peak structure is clearly
visible at $\gamma = -0.1$ in $L=8$, while at $\gamma = -0.4$ one has
to go to $L=12$ to observe an equivalent signal.

In what concerns the behavior of the free energy gap $\Delta F(L)$,
it grows for all investigated lattice sizes at all negative $\gamma$ values
(see Figure \ref{GAP_LAT} lower window). 
The value of $L$ at which $\Delta F(L)$
starts growing is certainly larger as the value of $\gamma$ is more negative.
This is another test of the increasing weakness of the transition as
$\gamma$ gets more negative. 

The statistics performed on the torus are reported in 
Table \ref{TABLA_TORO}. We also quote for the different 
negative $\gamma$, 
the value of $\beta^{\ast}(L)$ and the position of partition function
zero closest to the real axis. We have computed
from the imaginary part of the zeroes the effective
$\nu$ exponent between consecutive lattice sizes following (\ref{NU}).

In Figure \ref{NU_TORO} we plot for the different $\gamma$ values
the $\nu_{\rm {eff}}$ we measure. In all cases a $\nu_{\rm {eff}} \sim 1/3$
is observed for small lattice sizes, which gets closer to 0.25 when
the lattice is large enough. From this figure the trend of
$\nu_{\rm {eff}}$ seems rather clear towards the first order value.

From the energy distributions we measure the latent heat through a
cubic spline at the peaks.
Taking into account the value $\nu_{\rm {eff}}=0.25$ we measure,
we plotted it as a function of the inverse of the
volume $L^{-4}$, which is also the expected behavior of the latent
heat when the transition is first order.
The latent heat can be extrapolated to a value which
is safely far from zero (see Figure \ref{GAP_LAT} upper window).

In Figure \ref{PEAKS_TORO} we plot $C_v^{\rm {max}}(L)$ as a function
of $N_{\rm p}$. As could be expected from the behavior of the effective
exponent $\nu$, the maximum of the specific heat for small lattices diverges
slower than the volume. For the smaller lattices the effective 
exponent is $\alpha/\nu \sim 1.4$. This value increases monotonically
with the lattice size. In the largest ones we observe $\alpha/\nu \sim 3.5$.

\subsubsection{Spherical topology}

On the toroidal topology we have found that the minimum lattice
size required to observe a two state signal is obtained through
an apropriate combination $(\gamma,L_{\rm {min}})$, 
with increasing $L_{\rm {min}}$ for increasingly negative $\gamma$,
the behavior being qualitatively similar for all the $\gamma$ values
we have investigated. In view of this, we have studied a single
$\gamma < 0$ value on the spherical lattice to check if the
two state signal is absent on this topology. For the sake of comparison
with \cite{CN,JCN} we choose this value to be $\gamma = -0.2$.

\begin{figure}[h]
\begin{center}
\epsfig{figure= 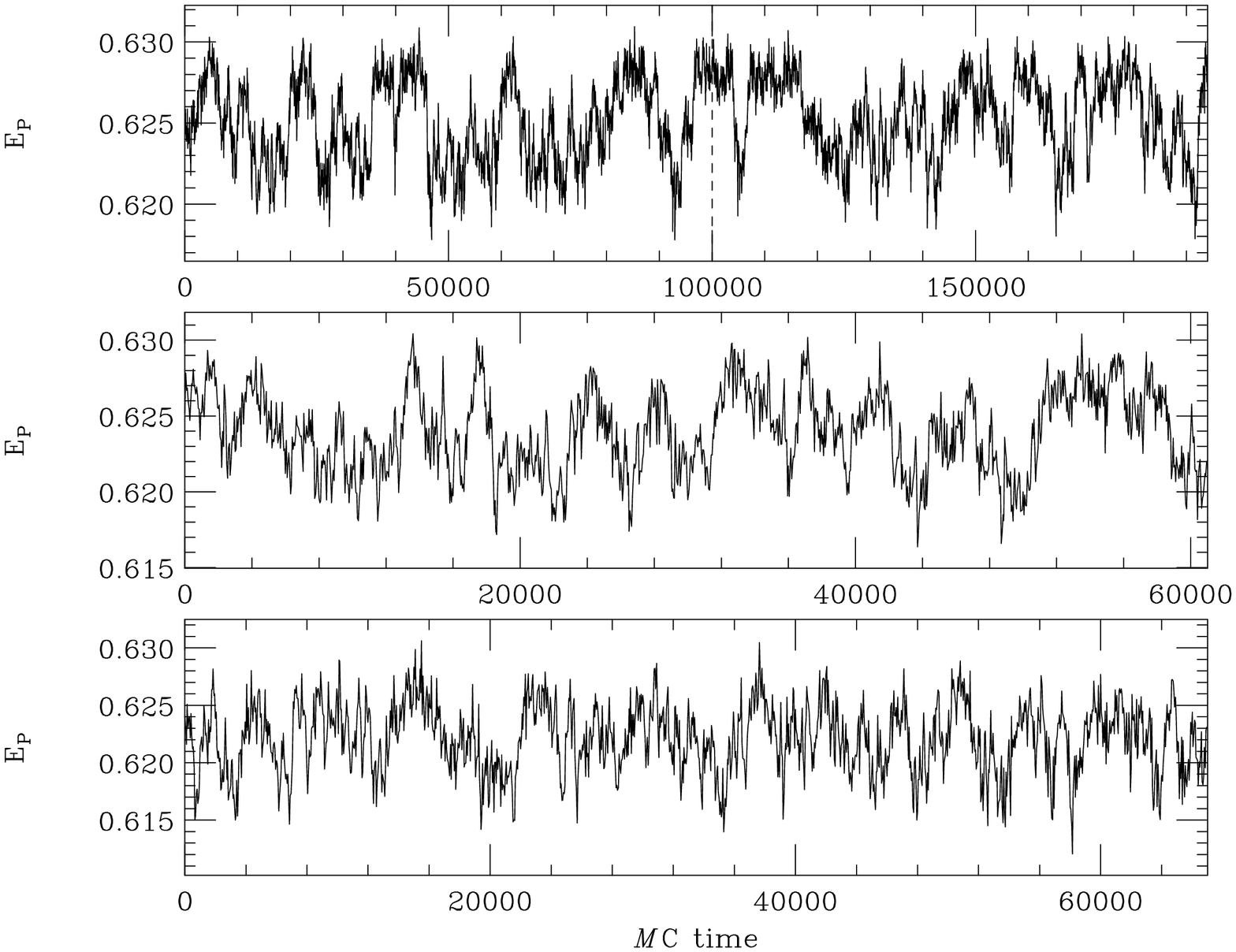,
angle=90,width=350pt}
\caption{\small {MC evolution of $E_{\rm p}$ in $N=12$ (lower window), $N=14$ 
(middle window) and $N=16$ (upper window) at $\gamma = -0.2$ on 
the $\HS$ topology. The two different runs in $N=16$ are separated by
a vertical dashed line.}}
\label{EVOL_02}
\end{center}
\end{figure}

We have run simulations on spheres ranging from $N=6$ ($L_{\rm {eff}} \sim 8$)
to $N=16$ ($L_{\rm {eff}} \sim 26$). The MC evolution is plotted in 
Figure \ref{EVOL_02}.

\begin{figure}[h]
\begin{center}
\epsfig{figure= 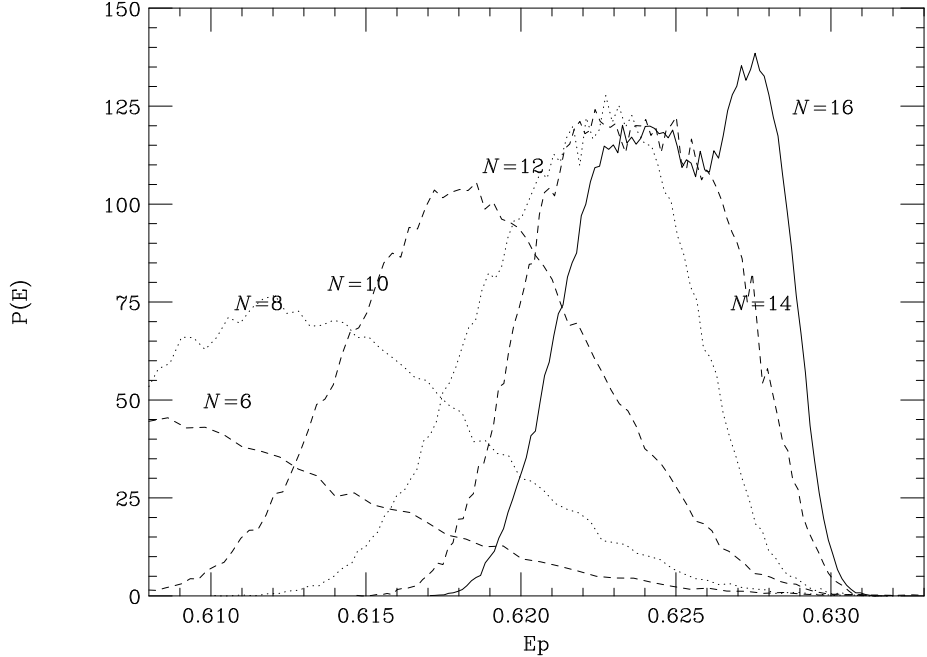,angle=90,width=350pt}
\caption{\small {$E_{\rm p}$ distributions at $\gamma = -0.2$ on 
the $\HS$ topology.}}
\label{HISTO_02}
\end{center}
\end{figure}

The distribution in $N=14$ is distinctly
non gaussian and the splitting of the peaks occurs in $N=16$ (see Figure
\ref{HISTO_02}). We remark that the simulations in $N=16$ are extremely
expensive in CPU. In order to alleviate thermalization we have parallelized
the code using shared memory in two PPro processors. In this lattice we
have run two independent simulations starting from different configurations
the results being fully compatible.

\begin{figure}[h]
\begin{center}
\epsfig{figure= 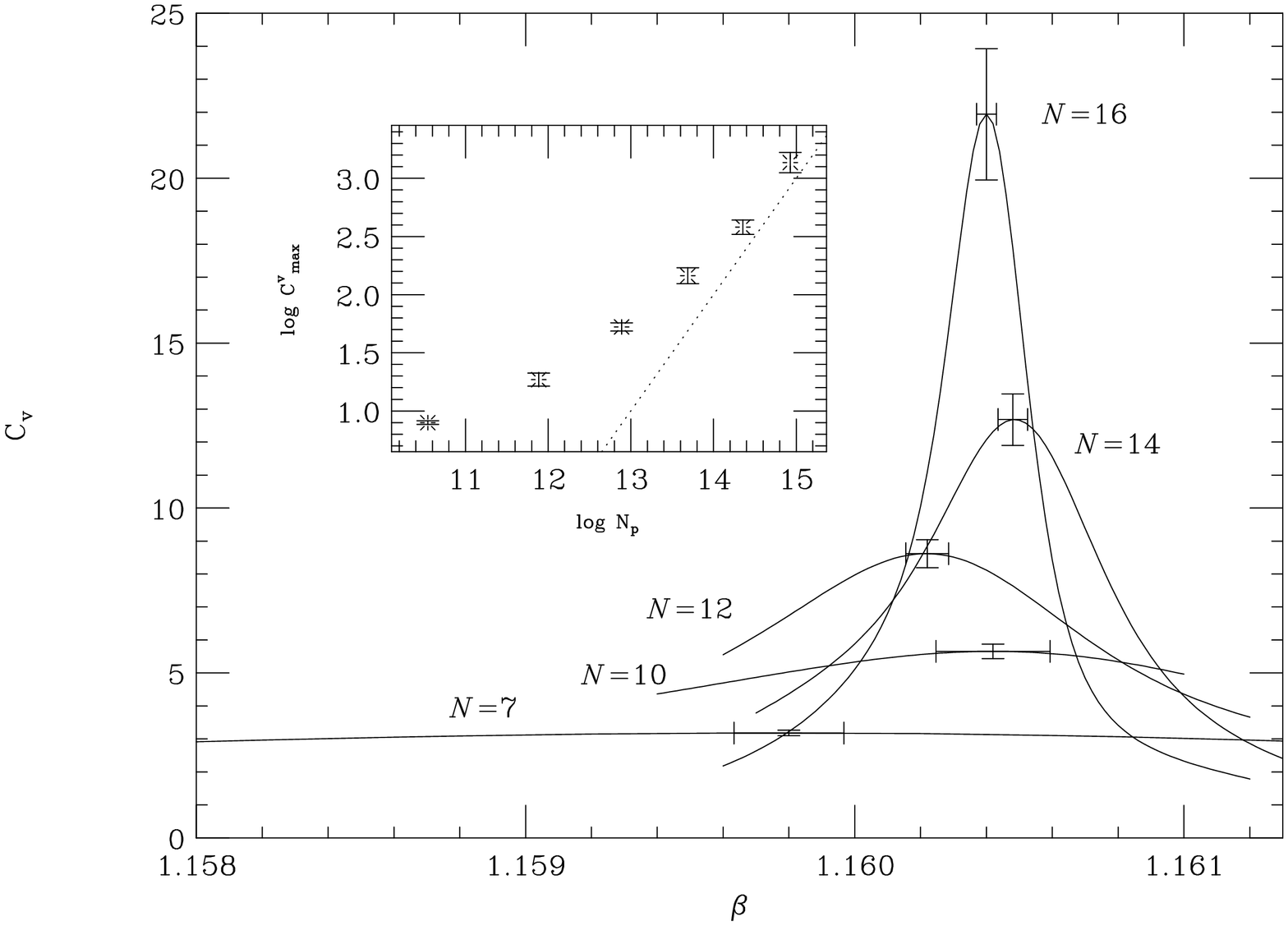,angle=90,width=350pt}
\caption{\small {Specific heat maximum, and Ferrenberg-Swendsen extrapolation 
(solid line) at $\gamma = -0.2$. 
The small window represents $C_{\rm v}^{\rm {max}} (N_{\rm p})$ 
The dotted line correspond to the behavior expected in a
first order phase transition.}}
\label{PEAKS_02}
\end{center}
\end{figure}

\begin{figure}[h]
\begin{center}
\epsfig{figure= 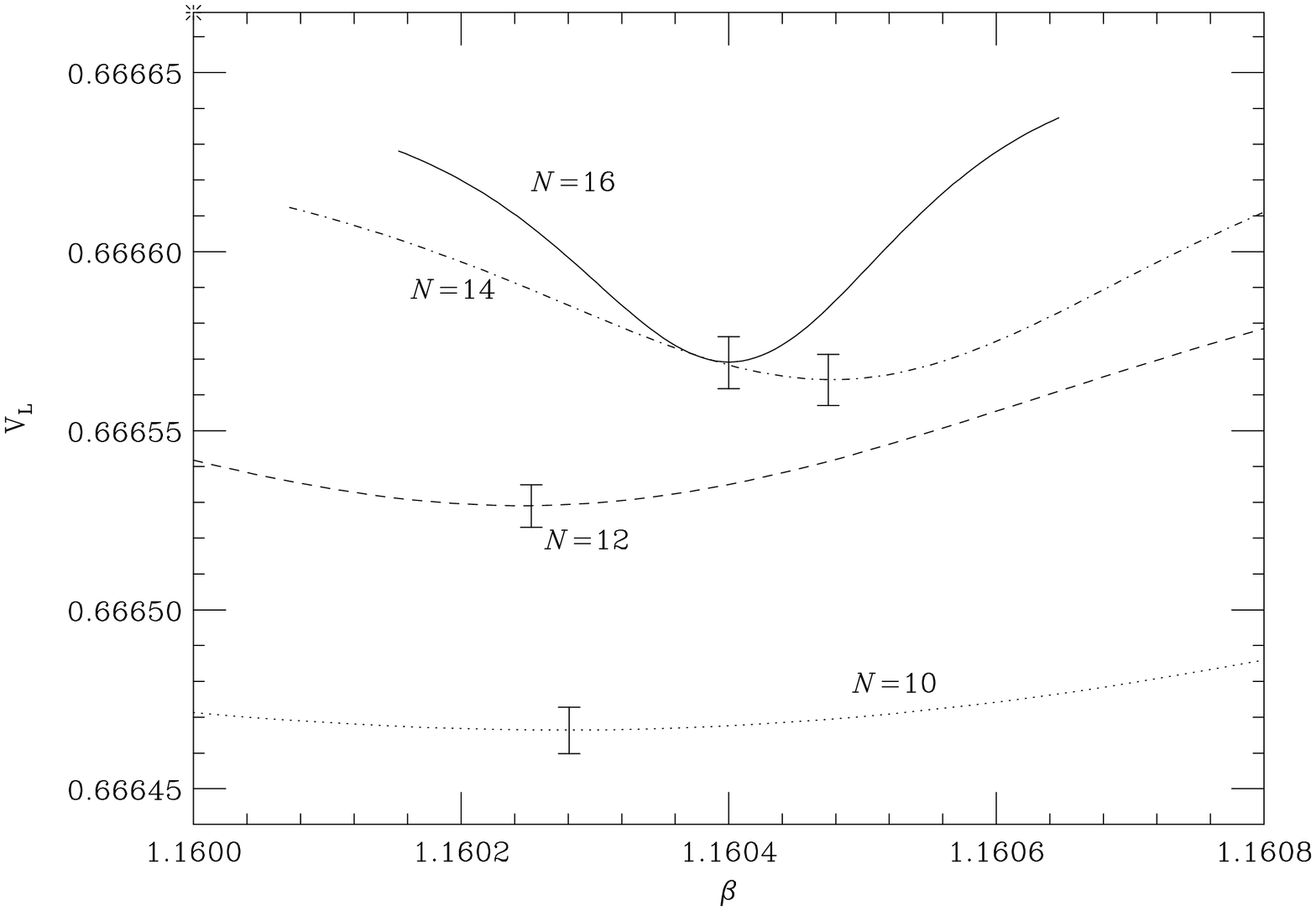,angle=90,width=350pt}
\caption{\small {Binder cumulant at $\gamma = -0.2$
on the $\HS$ topology in $N=10,12,14,16$. The cross in the upper corner 
signals the second order value $2/3$.}}
\label{BINDER_02}
\end{center}
\end{figure}

The behavior of the maximum of the specific heat is shown in Figure 
\ref{PEAKS_02}. We observe the same trend than in the previous values
of $\gamma$ in the sphere, i.e., an increasingly fast divergence
of the specific heat with the lattice size, showing an effective
$\alpha/\nu \sim 4$ already when the histogram width becomes constant 
with increasing lattice size. 

At this value of $\gamma$ we have a worse estimation for the latent
heat to be, as the two peaks have not split enough to allow an accurate
measurement.

Again the behavior of the Binder cumulant is very significant 
(see Figure \ref{BINDER_02}. The
value $V_L^{\rm {min}}$ shows a very fast trend towards 2/3 in the small
lattices. When increasing the lattice size the rate gets slower, and finally
the value in $N=16$ is compatible with the value in $N=14$ preempting
the extrapolation to $2/3$. Unfortunately lattices larger than $N=16$
are unaccessible to our computers nowadays, and we cannot observe
a decreasing $V_L^{\rm {min}}$ as we did for the other $\gamma$ values.
However from the behavior between $N=14$ and $N=16$ an increasing
$V_L^{\rm {min}}$ for larger spheres seems to be very unlikely.

Concerning the effective critical exponent $\nu$ we have measured
the position of the first Fisher zero (see Table \ref{TABLA_ESFERA})
and compute $\nu_{\rm {eff}}$ (see Figure \ref{NUESFERA}). Again the
first order value $1/d$ is reached in the largest lattices.

\subsection{Toroidal versus Spherical topology}

\begin{figure}[h]
\begin{center}
\epsfig{figure= 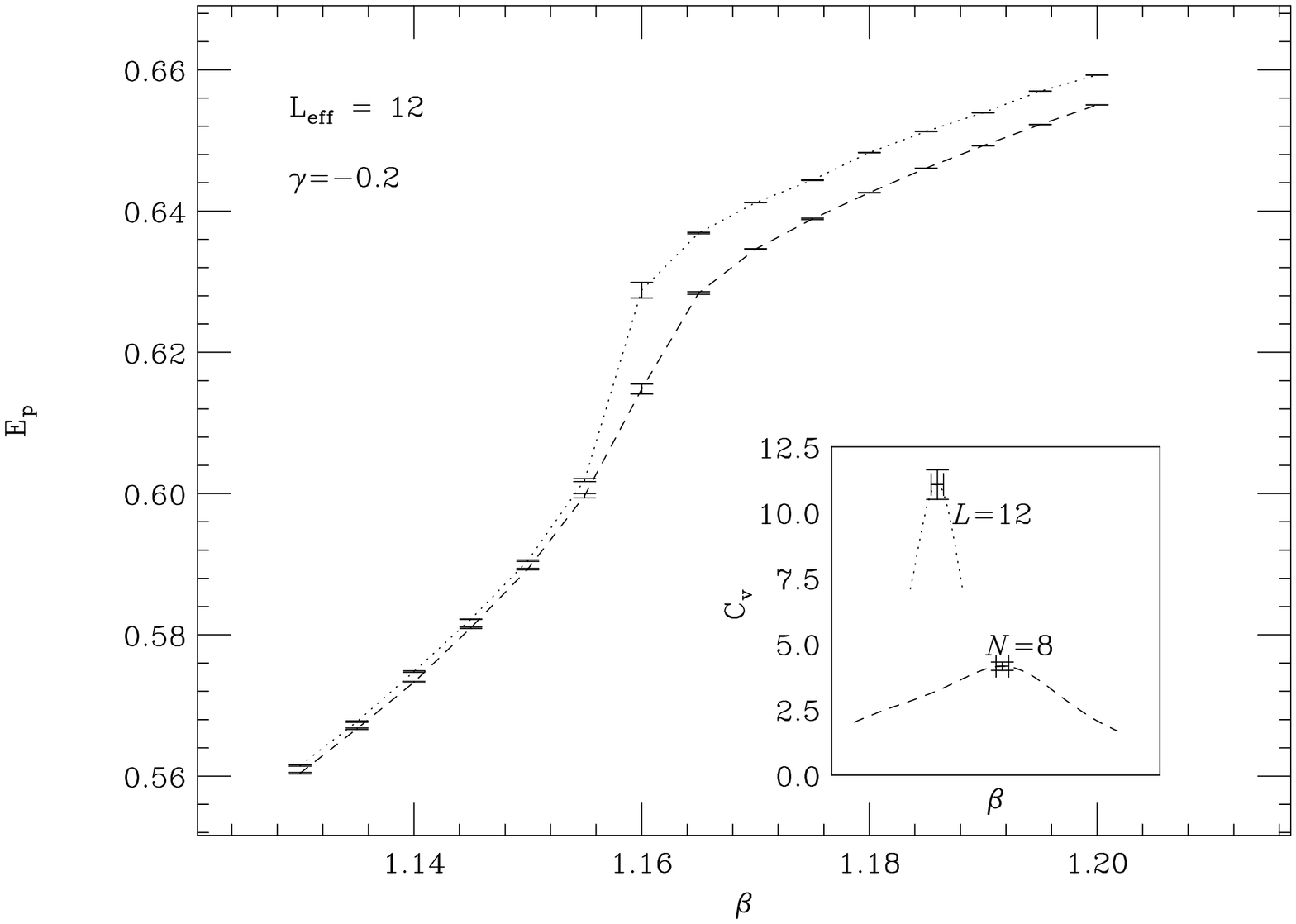,angle=90,width=350pt}
\caption{\small {$E_{\rm p}$ at $\gamma=-0.2$ in 
$L_{\rm {eff}} \sim 12$. The 
dotted line corresponds to the toroidal topology, the dashed one to the 
sphere $N=8$. The small window shows the difference in the 
specific heat between both cases.}}
\label{SALTO}
\end{center}
\end{figure}

\begin{figure}[h]
\begin{center}
\epsfig{figure= 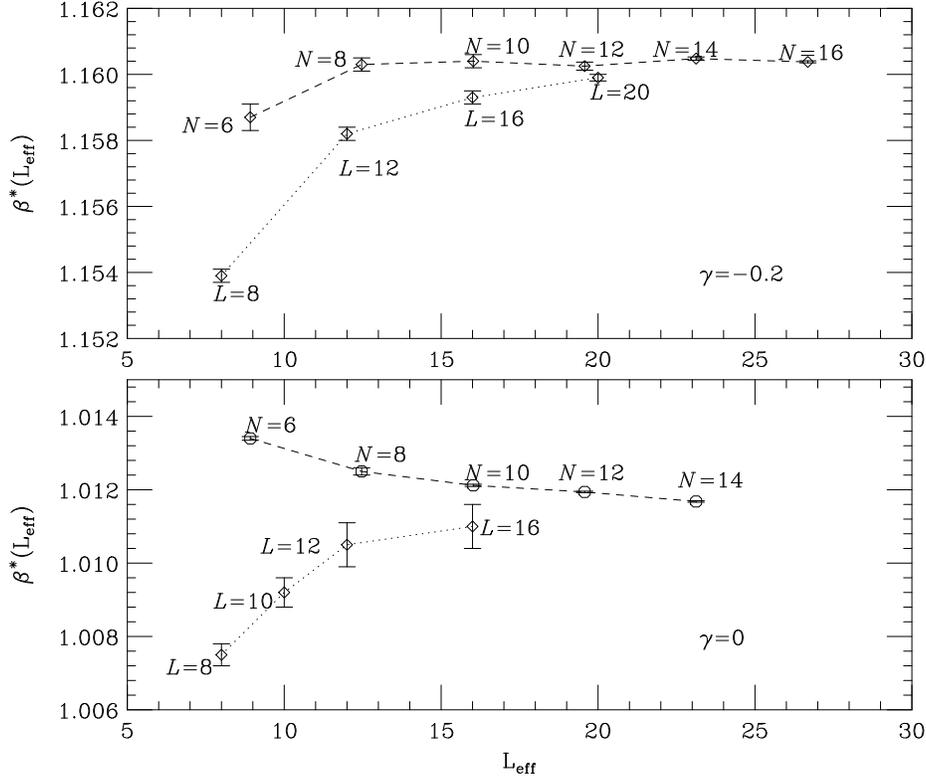,angle=90,width=350pt}
\caption{\small
{$\beta^{\ast}(L)$ for the different lattice sizes on the torus (dotted lines) 
and on the sphere (dashed lines) at $\gamma=0$ 
(lower window) and at $\gamma = -0.2$ 
(upper window). The couplings for $\gamma = 0$ on the torus have been taken
from \protect\cite{ALF}.}}
\label{BETAC}
\end{center}
\end{figure}

\begin{figure}[!h]
\begin{center}
\epsfig{figure= 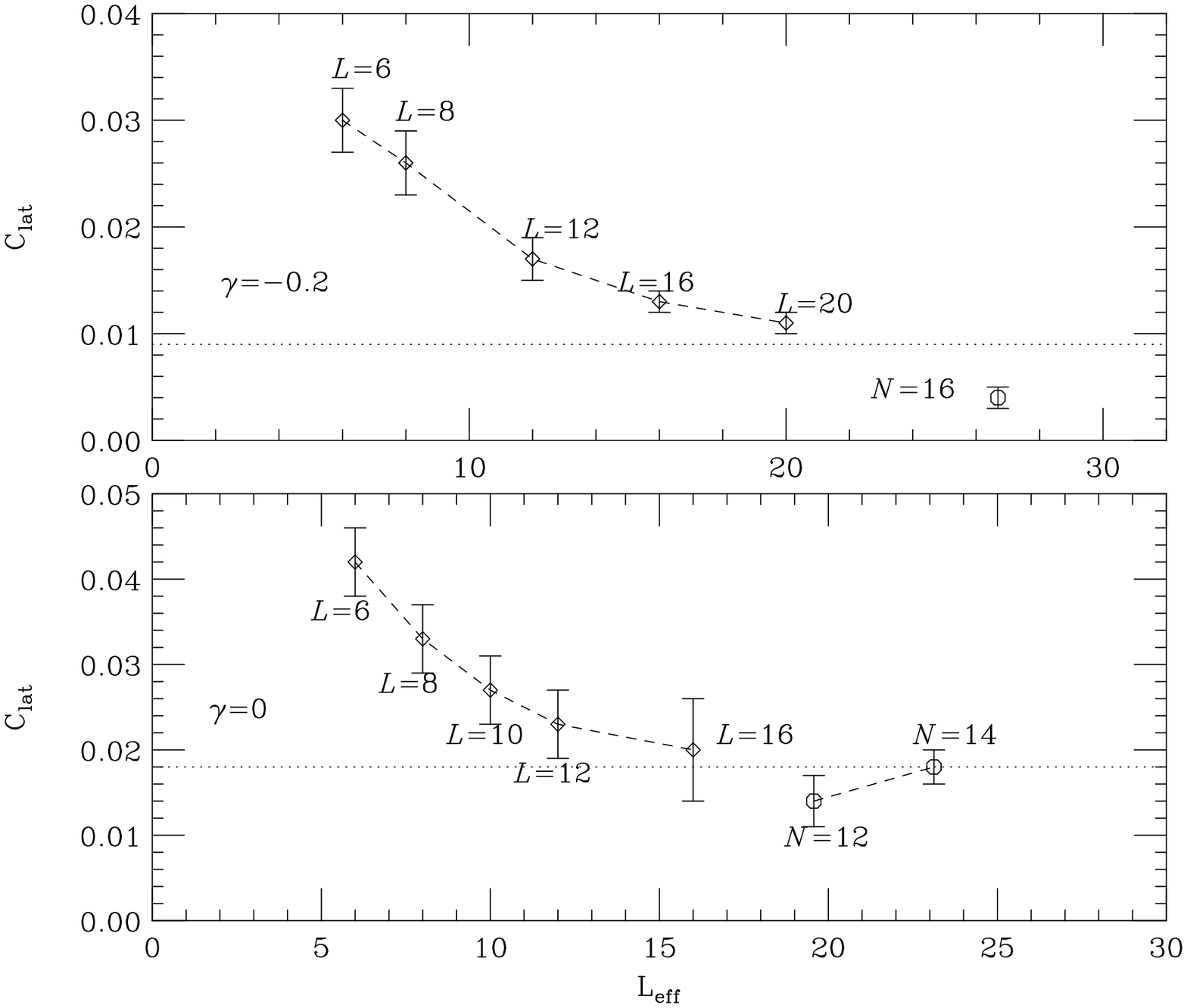,angle=90,width=350pt}
\caption{\small 
{Latent heat for the different lattice sizes on the torus and on the
sphere at $\gamma=0$ (lower window) and at $\gamma = -0.2$ (upper window).
The dotted line corresponds to $C_{\rm {lat}}(\infty)$ 
obtained by extrapolating
the values on the torus through a linear fit as function of $1/V$.
The values for $\gamma = 0$ on the torus have been taken
from \protect\cite{ALF}.}}
\label{LATCOMPA}
\end{center}
\end{figure}

Due to the fact that on the sphere there are a number of points
with less than the maximum connectivity, violations to standard
FSS in the form of uncontrolled finite size effects are expected
to appear on this topology. 

We have observed all along this work that working on the spherical
topology retards the onset of two-state signals.

In terms of $L_{\rm {eff}}$, 
some qualitative prediction arising from our results would be that
at fixed $\gamma$, the minimum lattice
size required to observe a double peak structure, $L_{\rm {min}}$, is
around three times larger on the sphere than on the torus.
This one being the showier difference, is not certainly the only one.

When comparing $C_{\rm v}^{\rm {max}}(L_{\rm {eff}})$
on both topologies one finds always smaller values on the spherical
topology (see small window in Figure \ref{SALTO}).
We did run on $L=12$ on the torus and $N=8$ on the sphere, which
has an $L_{\rm {eff}} \sim 12$, sweeping an interval
of $\beta$ values including the phase transition. In Figure \ref{SALTO}
we have plotted $E_{\rm p}$ in both topologies.
In the region of low $\beta$ the system is disordered,
the entropy is higher and the system is
not so sensitive to inhomogeneities of the lattice. However, in the
region of high $\beta$ the energy is in general smaller than for the
homogeneous lattice since the system tends to be ordered, and the influence
of the sites with less than maximum connectivity is more evident.
This difference gets smaller when the lattice size is increased on the
sphere, and hence the energy jump is larger. This feature explains
the observed splitting of the two states on the spherical
lattices when increasing lattice size.

Altogether, it is not recommended to work on spherical lattices to check
the existence or not of two-state signals. However, there are a number
of facts that make this topology not so dissapointing in spite of those
uncontrolled finite size effects.
The first one concerns the behavior of $\beta^{\ast}(L)$ and the second
one the measure of the latent heat.

In Figure \ref{BETAC} we compare $\beta^{\ast}(L)$ on the torus
and on the sphere. A first observation is that despite they not
having the same values in finite lattices, both curves get 
closer when increasing the lattice size. This supports the idea
of a common thermodynamic limit for both topologies.
On the other hand, the shift in the apparent critical coupling with $L$ is much
less dramatic on the sphere than on the torus. It seems that finite size
corrections to $\betac$ are smaller in the case of the spherical
topology. This behavior has also been observed in the Ising model
on spherical lattices \cite{JESUS}.

The behavior of the latent heat
is very significant. In Figure \ref{LATCOMPA} we compare the 
latent heat in finite volumes for both topologies. On the Wilson line,
at $\gamma = 0$ the asymptotic value of the latent heat is obtained
in the spherical topology. The inhomogeneity of the sphere has,
paradoxically, helped us to run lattices with $L_{\rm {eff}} \sim 24$
without having to worry about the tunneling rate. From this graph we
quote $C_{\rm {lat}} (\infty) = 0.018(1)$.
In the upper window the same is plotted for $\gamma = -0.2$. Unfortunately
we have just a single lattice size on the sphere to measure the latent
heat, however, from the behavior exhibited by this topology, 
an increase of the latent heat on spheres larger than $N=16$
seems rather likely.
Altogether, at $\gamma = -0.2$ a $C_{\rm {lat}} (\infty) \sim 0.009$
is plausible.

In what concerns the latent heat,
the spherical topology seems to afford an useful complement to the results
obtained for the the toroidal lattice. In fact, on the Wilson line
lattices larger than $L=16$ cannot be studied due to the technical problem
associated to the scarce tunneling. The spherical lattice
alleviates this technical problem, and the results
are supporting the value of the extrapolated latent heat as a 
function of the inverse of the volume, from the data on toroidal 
lattices up to $L=16$. One could consider the possibility of
simulating on spherical
lattices at negative $\gamma$  to solve the tunneling problem, but
in our opinion, the price to pay is too high.

\section{Conclusions}

The first order character of the deconfinement transition in pure
U(1) has been proved, up to the limits of a rather reasonable
numerical evidence, in the interval $\gamma \in [+0.2,-0.4]$.

In $\gamma = +0.2$ we have been able to stabilize the latent heat.
We are aware of no simulation on the torus showing a stable latent
heat due to the scarce tunneling in lattices larger than $L=6$.
The spherical topology has helped to solve this problem. However,
probably any lattice with inhomogeneities would produce the same
catalysing effects.

In $\gamma=0$ we have also been able to measure the 
asymptotic value of the latent heat.
We have proved the suggestions of several authors about a quasi
stabilization on lattices larger than $L \sim 12$. The data on the
spherical topology have been crucial to discard the possibility
of a slowly vanishing latent heat. 
It follows that the discretization of pure compact U(1) LGT on
the lattice using the Wilson action exhibits a first order phase
transition with a latent heat in the thermodynamical limit
$C_{\rm {lat}} \sim 0.018$.

On the toroidal topology things happen qualitatively in the same
way than in $\gamma=0$ up to $\gamma = -0.4$. We have run on spherical
lattices in $\gamma = -0.2$ looking for an argument similar to the one
found in the Wilson case concerning the stabilization of the latent heat.
The simulation in $N=16$ gives an estimation for the latent heat though
rather imprecise because the splitting of the peaks is not good
enough for the measurement to be accurate. However, in view of the
behavior exhibited by the spherical topology concerning
the trend of the two
peaks on spherical lattices to separate, we are prone to consider the
value measured in $N=16$ as a lower bound for the latent heat
in $\gamma=-0.2$. On the other hand,
we would at present say that the latent heat extrapolated
from the data on the torus up to $L=20$ is rather accurate, in view
of the behavior for the Wilson case.
The possibility of running larger spheres surpasses our computer
resources. A highly parallelized version of the code should
be used to alleviate thermalization, which is a possibility
we do not discard completely at medium term.

Either proving or discarding the possibility of a TCP at some
finite negative $| \gamma_{\rm {TCP}} | > 0.4$ will be a very
difficult task from the numerical point of view. An analytical
argument would be welcome.
A very tiny two-state signal, comparable with the one observed in $L=8$ at
$\gamma = -0.4$, is observed on the torus in $L=16$ at $\gamma = -0.8$,
which is the most negative value we have run on the torus.
Taking into account the factor 3 in $L_{\rm {eff}}$ one would
need a $N \sim 30$ sphere to observe a tiny double peak structure.
Simulation with spherical lattices in this range of $\gamma$ have no sense,
and nothing can be concluded from the absence of two state signals
from such negative $\gamma$ values.

In view of these difficulties to stabilize the latent heat for very
negative $\gamma$ values, the only chance to discern the order
of the phase transition is the study of effective critical exponents.
Cumulants of the energy, such as the Binder cumulant, have been
shown to behave like expected in the first order case
only when the stable two-peak structure is almost setting in, and there,
we do not need further evidences any longer. 
We could not expect additional information since by definition
the Binder cumulant
relies on the existence of stable latent heat in order to extrapolate
to $V_{\rm {L = \infty}}^{\rm {min}} < 2/3$. 

The advantage of studying the effective exponents (which is nothing
but studying the evolution of the histograms width) is that we do not
need a direct observation of latent heat to conclude that a transition
is first order \cite{SU2,ONAF}.
We have observed a $\nu_{\rm {eff}}$  which evolves
monotonically until it reaches the first order value, 
namely 0.25, in all cases.
The statistics needed to observe a monotonous behavior, 
are order $1000 \tau$ at the coupling $\beta^{\ast}(L)$, which has to be
located with high precision (four digits in our experience)
in order to accurately measure effective
exponents. 
However, one has to make every effort to observe the trend of the
effective exponents, since it is the only chance to discern the
order of such tricky transitions.

In pure U(1), for the Wilson case, an exponent $\nu \sim 1/3$ was widely
predicted \cite{LAU,BHA,BHA2} at the beginning of the eighties
when the lattice sizes where too small to reveal two peaks.
That $\nu$ was shown to become $\nu \sim 0.29$ when simulating
$L=14$ \cite{VIC2}. We have measured $\nu = 0.25$ on spherical lattices
and stated its first order character.
Within the approximation of effective potentials
it can be shown that along the transient region 
of a weak first order phase transition, everything goes
like in a second order one with a thermal exponent
$\alpha = 0.5$ \cite{LAF}. 
Together with Josephson law ($\alpha = 2 - \nu d$)
it implies $\nu \sim 0.37$. This statement has been checked in
2D Potts \cite{LAF}, 3D and 4D $O(N)$ models \cite{ISAF,ONAF,O4AF,O3AF}
and in the 4D SU(2)-Higgs at T=0 with fixed Higgs modulus \cite{SU2}.
Our results prove that pure compact U(1) theory behaves
in the same way.

A general remark to draw from this paper is though
there is no theorem constraining transitions presenting
$\nu_{\rm {eff}} \sim 1/3$ to be of WFO, yet the occurrence
of such exponent is a warning to consider the phase transition
as suspicious of being first order.

\section*{Acknowledgments} 
We have benefited from comments and discussions with
L.A. Fern\'andez and J. Salas. 
The discussions with C.B. Lang and J. Jers\'ak at the final
stage of the work have been very appreciated by the authors.

I.C. feels indebted to J. Jers\'ak for his many stimulating comments
as well as to C.B. Lang for his warm hospitality.

The simulations have been done using the {\sl RTNN} machine at 
Theoretical Physics Department, University of Zaragoza (Spain). 
I.C. is a MEC fellow.

\newpage

\section*{Appendix: Schwinger-Dyson Equations}

As a further check we have implemented the Schwinger-Dyson equations (SDE)
\cite{FALCI} which allow one to recover the simulated couplings 
from the Montecarlo data.

Let {\rm A($\theta$)} be an operator with null expectation value:
\begin{equation}
\langle A(\theta) \rangle = 
Z^{-1} \int [d\theta] A(\theta) e^{-S[\theta]} \equiv 0  \ .
\end{equation}

Derivating with respect to $\theta$ this equation trivially yields

\begin{equation}
\langle \frac{\partial A(\theta)}{\partial \theta} \rangle =
\langle A(\theta) \frac{\partial S[\theta]}{\partial \theta} \rangle \ .
\label{SD}
\end{equation}

which is the equation of movement of the operator {\rm A($\theta$)}
or Schwinger-Dyson equation. When the action depends on several 
couplings the equation (\ref{SD}) can be expressed as:

\begin{equation}
\langle \frac{\partial A(\theta)}{\partial \theta} \rangle =
\sum_i \beta_i \langle A(\theta) 
\frac{\partial S_i[\theta]}{\partial \theta} \rangle  \ .
\label{SDD}
\end{equation}

This equation relates the values of the couplings with the expectation
values we measure from the MC simulation. We need as many independent equations
as couplings we have to determine in the action. In our case, in order
to measure both $\beta$ and $\gamma$, we need two operators with null
expectation value in order to have two independent tests.
At each lattice site $n$ and for every direction $\mu$
we consider the operators:
\begin{eqnarray}
A(\theta) = \sin \theta_p = \sin(\theta_{n,\mu} - \theta_{stap}) \ , \\ 
B(\theta) = \sin 2\theta_p = \sin2(\theta_{n,\mu} - \theta_{stap}) \ ,
\end{eqnarray}

where $\theta_{stap}$ is the staple of the link labeled by ($n$,$\mu$).

Applying equation (\ref{SDD}) to those operators we get:
\small {
\begin{eqnarray}
\langle \sum_{\rm p} \cos(\theta_{n,\mu} - \theta_{stap}) \rangle = 
\beta \langle \sum_{\rm p} \sin(\theta_{n,\mu} - \theta_{stap})
\sum_{\rm p} \sin(\theta_{n,\mu} - \theta_{stap}) \rangle + \nonumber  \\ 
 + 2\gamma \langle \sum_{\rm p} \sin(\theta_{n,\mu} - \theta_{stap}) 
\sum_{\rm p} \sin2(\theta_{n,\mu} - \theta_{stap}) \rangle 
\end{eqnarray}

\begin{eqnarray}
\langle \sum_{\rm p} 2\cos2(\theta_{n,\mu} - \theta_{stap}) \rangle = 
\beta \langle \sum_{\rm p} \sin2(\theta_{n,\mu} - \theta_{stap})
\sum_{\rm p} \sin(\theta_{n,\mu} - \theta_{stap}) \rangle + \nonumber \\ 
 + 2\gamma \langle \sum_{\rm p} \sin2(\theta_{n,\mu} - \theta_{stap})
\sum_{\rm p} \sin2(\theta_{n,\mu} - \theta_{stap}) \rangle 
\end{eqnarray}
}

On the hypertorus $\sum_{\rm p}$ means the sum over the plaquettes
in positive and negative directions ($\pm \mu$)
bordering the link (12 plaquettes). On the sphere one has to be careful,
since not all the links have 12 surrounding plaquettes, and the sum
has to be understood as extended to the existing plaquettes.

These equations hold for all $n$ 
in such a way that we can sum up the equations
for every single link and average over the number of links.
We quote in Table \ref{TABLA_SD} for the tests done at $\gamma = -0.2$
using both topologies. They show a perfect agreement between the simulated
couplings and the ones recovered from the MC expectation values.

\begin{table}[h]
{

\begin{center}
{
\small{
\begin{tabular}{|c||c|c||c||c|c|c|} \hline
\multicolumn{3}{|c||}{$\HT$ topology} &  \multicolumn{3}{|c|}{$\HS$ topology} \\ \hline  \hline
$L$ & $\beta_{\rm {sim}}$ & $(\beta,\gamma)_{\rm {SD}}$ 
&N &$\beta_{\rm {sim}}$ &$(\beta,\gamma)_{\rm {SD}}$   \\ \hline
$6$ &1.1460 &1.1466(44),-0.2009(26)        
&6 &1.1587 &1.1592(14),-0.1998(12)                     \\ \hline
$8$ &1.1535  &1.1527(22) -0.1997(15)       
&8  &1.1600  &1.1599(12) -0.1999(8)               \\ \hline
$12$ &1.1582 &1.1581(12),-0.1999(8)       
&10  &1.1602 &1.1602(7),-0.2001(9)                   \\ \hline
\hline

\end{tabular}
}
}
\end{center}

}
\caption[a]{\small{Couplings obtained from the MC simulations at $\gamma$=-0.2
using the Schwinger-Dyson equations.}}
\protect\label{TABLA_SD}

\end{table}

\end{document}